\documentclass[amsmath,amssymb,aps]{revtex4-2}

\usepackage{graphicx}
\usepackage{dcolumn}
\usepackage{footmisc}
\usepackage{natbib}

\usepackage{bm}
\usepackage{hyperref}
\usepackage{xcolor}
\usepackage{float}

\usepackage{scalerel}
\usepackage{tikz}
\usetikzlibrary{svg.path}

\definecolor{orcidlogocol}{HTML}{A6CE39}
\tikzset{
  orcidlogo/.pic={
    \fill[orcidlogocol] svg{M256,128c0,70.7-57.3,128-128,128C57.3,256,0,198.7,0,128C0,57.3,57.3,0,128,0C198.7,0,256,57.3,256,128z};
    \fill[white] svg{M86.3,186.2H70.9V79.1h15.4v48.4V186.2z}
                 svg{M108.9,79.1h41.6c39.6,0,57,28.3,57,53.6c0,27.5-21.5,53.6-56.8,53.6h-41.8V79.1z M124.3,172.4h24.5c34.9,0,42.9-26.5,42.9-39.7c0-21.5-13.7-39.7-43.7-39.7h-23.7V172.4z}
                 svg{M88.7,56.8c0,5.5-4.5,10.1-10.1,10.1c-5.6,0-10.1-4.6-10.1-10.1c0-5.6,4.5-10.1,10.1-10.1C84.2,46.7,88.7,51.3,88.7,56.8z};
  }
}

\newcommand\orcidicon[1]{\href{https://orcid.org/#1}{\mbox{\scalerel*{
\begin{tikzpicture}[yscale=-1,transform shape]
\pic{orcidlogo};
\end{tikzpicture}
}{|}}}}

\begin{document}
\setlength{\abovedisplayskip}{5pt}
\setlength{\belowdisplayskip}{5pt}
\setlength{\abovedisplayshortskip}{5pt}
\setlength{\belowdisplayshortskip}{5pt}

\def\mnras{M.N.R.A.S.}

\def\ak#1{\textcolor{magenta}{{\bf AK}: #1}}
\def\ad#1{\textcolor{blue}{{\bf AD}: #1}}

\preprint{}

\title{Constraints on blue and red tilted primordial power spectra\\ using dwarf galaxy properties}

\author{Ariane Dekker$^{1, \orcidicon{0000-0002-3831-9442}}$ and Andrey Kravtsov$^{1, 2, 3\orcidicon{0000-0003-4307-634X}}$}
\email{ahdekker@uchicago.edu, kravtsov@uchicago.edu}
\affiliation{$^1$Kavli Institute for Cosmological Physics, The University of Chicago, Chicago, IL 60637 USA}
\affiliation{$^2$Department of Astronomy and Astrophysics, The University of Chicago, Chicago, IL 60637 USA}
\affiliation{$^3$Enrico Fermi Institute, The University of Chicago, Chicago, IL 60637}

\begin{abstract}
Although the standard $\Lambda$+Cold Dark Matter ($\Lambda$CDM) model is well tested on large scales, the primordial power spectrum may deviate from the $\Lambda$CDM spectrum on small scales due to specific dark matter properties or alternative inflationary models. 
These deviations affect the formation of dark matter structure, which subsequently leads to different observable properties of galaxies. 
In this work, we study the impact of a blue and red tilted power spectrum on the central density of dwarf galaxies. To do this, we model densities of dwarf galaxies using a combination of high-resolution numerical simulations and galaxy formation model. The model galaxies in $\Lambda$CDM are consistent with observations of 41 faint dwarf satellite galaxies of the Milky Way. The deviations from the $\Lambda$CDM power spectrum are constrained by the central matter densities of dwarf galaxies, which set stringent constraints on the possible small-scale tilt of the primordial power spectrum, improving on the current limits. Moreover, similar analysis can be applied to test any feature in the power spectrum at small scales between $k\sim 10-100$~Mpc$^{-1}$. 
\end{abstract}

\maketitle

\section{\label{intro} Introduction}
Structure in the Universe has formed through gravitational instabilities of the primordial matter density fluctuations. These fluctuations are thought to arise from the quantum fluctuations during inflation, which are stretched to astrophysical 
scales due to accelerated expansion \citep{Mukhanov.Chibisov.1981,Starobinsky.1982,Linde.1982,Hawking.1982}. In standard inflationary models, the dynamics are described by a homogeneous single scalar field, the inflaton, with a kinetic energy that slowly rolls on a potential (see Ref.~\cite{baumann2012tasi} for a review). Perturbations are constantly created and are predicted to be nearly scale-invariant, Gaussian, and adiabatic. 
The initial perturbations can be characterised in the Fourier space by the primordial matter power spectrum as a function of wavenumber $k=2\pi/l$ corresponding to a given spatial scale $l$. 

The initial perturbation spectrum predicted in the inflation and $\Lambda$CDM model has been confirmed by observations over large length scales between 0.003~Mpc$^{-1}\lesssim k \lesssim 3$~Mpc$^{-1}$ by the CMB~\cite{2020}, galaxy clustering~\cite{Troxel_2018,2017AJ....154...28B} and Lyman-$\alpha$ measurements~\cite{Chabanier_2019}. Within this range of scales, a nearly scale-invariant power spectrum is found, consistent with standard inflationary models. 

At smaller scales, the shape of the power spectrum remains relatively unconstrained and may deviate from the standard $\Lambda$ Cold Dark Matter ($\Lambda$CDM) model by showing a tilt, spikes, bumps, or suppression, as motivated by various inflation~\cite{Bartolo:2004if,Bridle_2003} and dark matter ~\cite{Bechtol:2022koa,Tulin:2017ara}.
In particular, an enhancement in density perturbations is expected in models with multiple fields during inflation, vector dark matter produced during inflation, axion production through misalignment mechanism, non-Gaussianity and scale-dependency~\cite{Padilla_2019,PhysRevD.88.103518,Biagetti_2015,Iarygina:2020dwe,Wands,Ebadi:2023xhq,PhysRevD.81.043534,PhysRevD.85.123537,Chung_2015,Talebian:2022jkb,Graham_2016,PhysRevD.86.103508,Bartolo_2016,inflation_2016,Cook_2012,Arvanitaki:2019rax}
Moreover, the presence of primordial magnetic fields enhances baryon perturbations, producing bumps in the matter power spectrum at small scales~\cite{Katz:2021iou,Ralegankar:2024ekl}. A suppression of density perturbation is predicted by various dark matter models with large free-streaming lengths or self-interactions~\cite{Lovell:2013ola,Boyarsky:2009ix,Egana-Ugrinovic:2021gnu}. 

Deviations from the spectrum in the standard $\Lambda$CDM model can lead to interesting features in the collapsed structures. An enhancement in power on small scales, for example, can result in the formation of primordial black holes (PBH), and a sufficiently large enhancement can potentially produce enough PBHs to account for dark matter~\cite{Yi_2023,Braglia_2020,Arya:2019wck}. 
Moreover, large primordial fluctuations can induce a stochastic gravitational wave background that is detectable with gravitational wave detectors~\cite{Domenech:2021ztg,Ebadi:2023xhq,Alabidi:2012ex}. Interestingly, the recent results of pulsar timing arrays by NANOGrav~\cite{NANOGrav:2023gor}, EPTA~\cite{EPTA:2023fyk}, CPTA~\cite{Xu:2023wog} and PPTA~\cite{Reardon:2023gzh} find evidence for a stochastic gravitational wave background, which can currently be interpreted by cosmological models that enhance primordial fluctuations, as well as in terms of inspiraling supermassive black hole binaries~\cite{Afzal_2023}. If these changes in the density perturbations have a primordial origin, they can be tested through spectral distortions in the CMB frequency spectrum, detectable by instruments such as COBE/FIRAS~\cite{Nakama:2017ohe,Kogut:2019vqh}. 

Probing the power spectrum on small scales is challenging because gravitational collapse that forms dark matter halos and galaxies is a nonlinear process, and the mapping between observations and the primordial power spectrum requires an accurate understanding of dark matter structure formation and galaxy formation physics. However, observations of the abundance, luminosities and velocity dispersions of nearby dwarf galaxies have successfully been used to constrain power spectra with a strong suppression on small scales, caused by free-streaming and self-interacting properties of dark matter models, such as the thermal relic warm dark matter, sterile neutrino, self-interacting dark matter or fuzzy dark matter~\cite{Kennedy_2014,DES:2020fxi,Nadler.etal.2021,kim2022milky,PhysRevD.95.083015,Newton:2020cog,Dekker_2022,Dalal.Kravtsov.2022}. 
Models that enhance the matter power spectrum are however less studied. Recently, two studies have set constraints on the matter power spectrum with a positive tilt. Thus, Ref. \cite{Gilman_2022} constrains the shape of the matter power spectrum using strong gravitational lensing to be close to the $\Lambda$CDM prediction on scales up to $k \leq 70$~Mpc$^{-1}$.   Ref.~\cite{esteban2023milky} performed a joint analysis of the internal velocity dispersion of stars, size, and total abundance of Milky Way satellite galaxies to constrain the matter power spectrum at 4~Mpc$^{-1} \leq k \leq 37$~Mpc$^{-1}$. The velocity dispersion and size allow to constrain the central matter density of galaxies. Given that the mass in faint dwarf galaxies is dominated by dark matter, the central density is related to the initial amplitude of the matter power spectrum. 

In this work, we extend the work of \cite{esteban2023milky} by using a larger sample of 41 Milky Way dwarf satellite galaxies with velocity dispersion measurements and projected half-light radii (compiled in the Appendix B of \cite{Kravtsov_2023}) to test power spectra with enhanced and suppressed density perturbations at small scales. Extending modeling to smaller galaxy masses allows us to extend power spectrum constraints to smaller scales. Furthermore, instead of modeling the luminosity-size and luminosity-halo mass relations using a parameterized relations, we use a galaxy formation model of \cite{2022MNRAS.514.2667K} that was shown to faithfully reproduce many observed properties of dwarf galaxies in $\Lambda$CDM.

An enhancement (suppression) of density perturbation leads however to an earlier on-set (delay) of dark matter halo formation, consequently enhancing (suppressing) dark matter halo concentration. We thus analyze the differences in the concentration of dark matter halos for models with different power spectra thereby setting constraints on the matter power spectrum at $z=0$, and interpret these in terms of the primordial power spectrum.

The paper is structured as follows. Section~\ref{sec2} describes the matter power spectrum model adopted, followed by a description of modeling the inner mass of galaxies. Section~\ref{sec3} describes the analysis, and section~\ref{sec4} presents the results and discussion. 

\section{\label{sec2} Modeling the inner mass of galaxies}

Observations of local dwarf galaxies show a correlation between the total mass within the projected stellar half-light radius $M_{\rm tot}(<r_{1/2})$ and the galaxy luminosity in the $V$-band $L_{\rm V}$~\cite{Simon:2019nxf,Kravtsov_2023,Kravtsov:2024esa}. The total mass $M_{\rm tot}(<r_{1/2})$ can be estimated accurately using galaxy line-of-sight velocity dispersion \cite{Wolf_2010}. Dark matter and stars both contribute to mass, but given that faint galaxies are dark matter-dominated, the mass within half-light radius is sensitive to the concentration of the dark matter halo. The latter depends on the formation time of dark matter halos, which, in turn, depends on the amplitude of primordial perturbations at the scale corresponding to the average size of the region that collapses into the halo hosting a given galaxy. 

Specifically, the Fourier transform of the Lagrangian region of average radius $R$ collapsing into a halo has an extent of $k\approx 4.5/R$ \citep{Chan:2015zjt}. At the same time, halo concentration, defined as the ratio between the halo virial radius and the inner scale radius $r_s$ where the slope of the density profile is -2, $c(M,z)=r_{200}(M,z)/r_s$, depends on the collapse of the main progenitor of a smaller mass at high redshift and mergers with halos of smaller mass. The effective scale determining halo concentration is $\approx 0.4R$ \cite{Diemer.Joyce.2019}. Given that $R= [3M/(4\pi \bar{\rho}_{\rm m})]^{1/3}$, the wavenumber of the power spectrum probed by halos of mass $M$ is 
\begin{equation}
\label{eq1}
k \approx \frac{4.5}{0.4}\,\left(\frac{4\pi\bar{\rho}_{\rm m}}{3M}\right)^{1/3}\approx 134.5\,{\rm Mpc^{-1}}\left(\frac{M}{10^8\,M_\odot}\right)^{-1/3}\left(\frac{\Omega_{\rm m0}}{0.31}\right)^{1/3}\left(\frac{H_0}{70}\right)^{2/3}.
\end{equation}

The estimates of $M_{\rm tot}(<r_{1/2})$ for the faintest dwarf galaxies, which are expected to be hosted by the smallest mass halos of $M\sim 10^8\, M_\odot$ are therefore a promising probe of the primordial power spectrum amplitude at scales up to $k\sim 100\,\rm Mpc^{-1}$. Indeed, in this section we discuss the dependence of the correlation between $M_{\rm tot}(<r_{1/2})$ and $L_{\rm V}$ on the amplitude of the primordial power spectrum at the relevant scales. We also describe our model for halo and galaxy formation, as well as observational measurements of the Milky Way dwarf satellite galaxies. 

\subsection{Primordial matter power spectrum}
We assume that the primordial power spectrum in $\Lambda$CDM is described by $P_{\rm{prim}} (k)=A_s \left(k/k_0\right)^{n_s}$, where $A_s=2.100\pm 0.030\times 10^{-9}$ is the amplitude of the power spectrum which is observed through the CMB anisotropy amplitude measured at scale $k_0=0.05$~Mpc$^{-1}$~\cite{2020}. The scalar spectral index $n_s(k)=d\ln P_{\rm{prim}}(k)/d\ln k$ varies very slowly with $|n_s-1| \ll 1$ in standard slow-roll inflationary models, and, indeed, analyses of CMB temperature fluctuations estimate $n_s=0.9649\pm 0.0042 $ \cite{2020}. 
In this work, we use the total masses within half-light radii of observed Milky Way dwarf satellites to constrain blue and red tilted matter power spectra, and discuss implications in section~\ref{sec4}. In particular, we consider a model-independent primordial power spectrum that scales as $k^{m_s}$ beyond a pivot scale $k_p$ as has been discussed in previous works~\cite{Parashari_2023,Hirano:2015wla}
\begin{align}
    P_{\rm{prim}}(k) &\propto k^{n_s} &(k\leq k_p)\\
    &\propto k^{n_s} \left(\frac{k}{k_p}\right)^{m_s-n_s} &(k> k_p), 
\end{align}
Here, $m_s>n_s$ corresponds to a blue tilted power spectrum, $m_s<n_s$ to a red tilted power spectrum, and $m_s\equiv n_s$ is the standard $\Lambda$CDM cosmological model. We assume that the perturbations evolve as in $\Lambda$CDM, and that the linear matter power spectrum today is related to the primordial power spectrum through a transfer function as $P_{\rm{lin}}\equiv P_{\rm{prim}}(k) T^2(k)$~\cite{1998ApJ...496..605E}. Throughout this work, we adopt the results from Planck 2018 with cosmological constants $h=67$, $\sigma_8=0.81$, $\Omega_{\Lambda}=0.69$, $\Omega_m=0.31$ and $\Omega_{b}=0.05$~\cite{2020}.

\subsection{Halo and galaxy formation models}

The formation time of dark matter halos and the concentration of their final density profile depend on the amplitude of the initial perturbations from which they collapse. The initial primordial perturbation spectrum with the blue (red) tilt beyond the pivot scale enhances (suppresses) the formation of structure at $k>k_p$. This affects both the abundance and concentration of dark matter halos of mass $M<M_p$, where $M_p\propto k^{-3}_p$.

\begin{figure*}[ht!]
    \centering
    {{\includegraphics[width=0.45\textwidth]{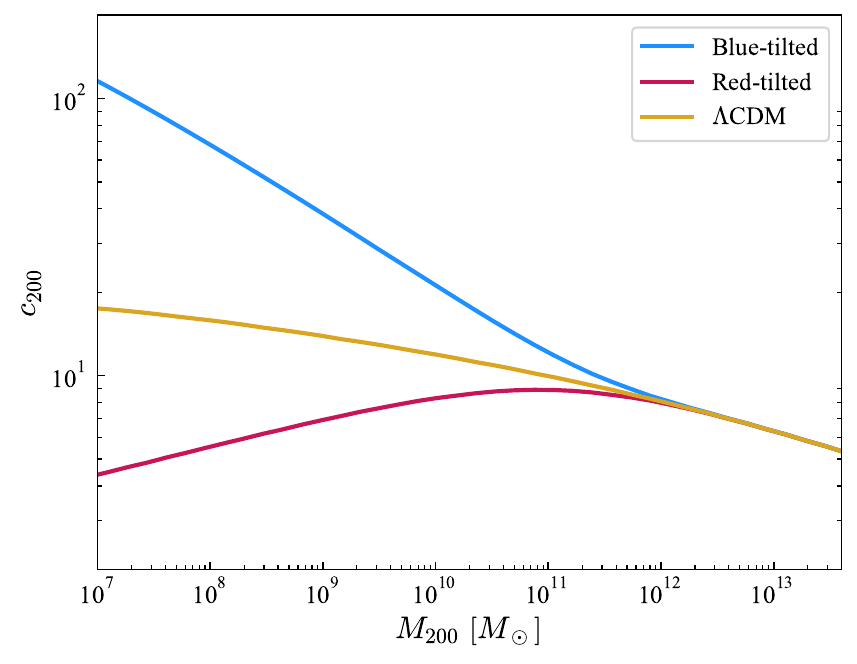} }}
 \qquad
    {{\includegraphics[width=0.45\textwidth]{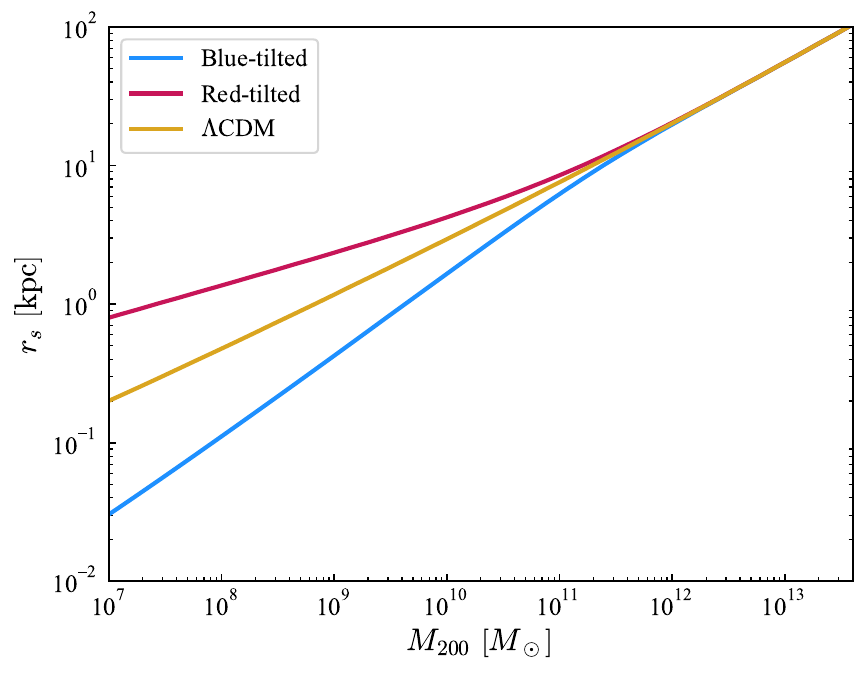} }}
\caption{Concentration (left) and scale radius (right) as a function of virial mass at $z=0$, assuming a blue-tilted ($m_s=2.0$, $k_p=5$~Mpc$^{-1}$), red-tilted ($m_s=0.0$, $k_p=5$~Mpc$^{-1}$) and $\Lambda$CDM matter power spectrum computed using the model of Ref.~\cite{Ludlow:2016ifl}.
}\label{fig1}
\end{figure*}

We use the concentration calibration of Ref.~\cite{Ludlow:2016ifl} to compute the concentration as a function of mass and redshift for an input primordial power spectrum. 
The concentration prediction of Ref.~\cite{Ludlow:2016ifl} has been tested for suppressed and enhanced primordial power spectra~\cite{Brown:2020lxj}. 
Figure~\ref{fig1} shows the concentration (left) and scale radius (right) assuming a blue-tilted ($m_s=2.0$, $k_p=5.0h$~Mpc$^{-1}$),  red-tilted ($m_s=0.0$, $k_p=5.0h$~Mpc$^{-1}$), and $\Lambda$CDM power spectrum at $z=0$. For benchmark values adopted for ($m_s,k_p$) it can be seen that the concentration and scale radius deviate from the standard $\Lambda$CDM for halo mass up to $\sim10^{12}$~M$_{\odot}$, which would give a distinct signature in the dwarf galaxy observations. 

We use a two-stage approach to modeling effects of the power spectrum on $M_{\rm tot}(<r_{1/2})$ of dwarf galaxies. 
To estimate these masses in the $\Lambda$CDM model, we use the results from the high-resolution Caterpillar simulation suite of 32 Milky-Way sized halos ~\cite{Griffen_2016}. Specifically, we use the evolution of mass of halos and subhalos extracted from the simulations and evolve properties of galaxies that are expected to form in each halo using a regulator-type galaxy formation model of \cite{2022MNRAS.514.2667K}. 
This framework follows galaxy evolution using a system of coupled differential equations that describe the evolution of the interstellar medium (ISM) and stellar masses of galaxies, as well as stellar mass and metallicity of the ISM and stars. The model also accounts for the effects of UV heating on the intergalactic medium after reionization and feedback-drive outflows. 

The model predicts half-light radius for each galaxy and we use the evolution of stellar mass and metallicity computed by the model and the Flexible Stellar Population Synthesis (FSPS) code~\cite{Conroy:2008mp,Conroy:2009vq} to obtain galaxy luminosity in the V-band. The fiducial $\Lambda$CDM model reproduces a wide range of observed dwarf galaxy properties such as the luminosity function, radial distribution, and size-luminosity and metallicity-mass relation of Milky Way satellites, as was demonstrated in Ref.~\citep[][see Table 1 in their work for fiducial values that we adopt for the gas inflow, reionization, gas disk, star formation, galactic outflow and chemical evolution models]{2022MNRAS.514.2667K}. 

The total mass within the half-light radii in the model is given by
\begin{equation}
    M_{\rm tot}(<r_{1/2})=M_{\rm dm, NFW}(<r_{1/2})+\frac{1}{2}M_{\star},
\end{equation}
where $r_{1/2}$ and $M_{\star}$ are predicted by the model, and $M_{\rm dm, NFW}(r)$ is the total mass within $r_{1/2}$, which we compute assuming the Navarro–Frenk–White density profile~\cite{Navarro:1996gj} and the scale radius estimated for each halo and subhalo in the simulation halo catalog. We do not include gas mass in this estimate because all of the MW satellites except for the Small and Large Magellanic Clouds have no ISM gas, which was likely stripped during the orbital evolution of satellites. 

To estimate the effect of changing power spectrum on $M_{\rm tot}(<r_{1/2})$, we assume that the main effect of the change is on the halo concentration, while galaxy evolution is assumed to be unaffected. Although some effect can be expected, it is likely to be small, while effect on the concentration is direct and substantial. Specifically, we rescale the scale radii of each halo from their $\Lambda$CDM simulation values using the ratio of concentrations in the model with a modified spectrum and the $\Lambda$CDM concentrations. 

\subsection{Sample of observed dwarf Milky Way satellites}

We use a sample of 41 Milky Way dwarf satellites with observational estimates of V-band luminosity, velocity dispersion, and projected half-light radius presented in Ref.~\cite{Kravtsov_2023}. These satellites have been discovered in various surveys such as the Sloan Digital Sky Survey, Dark Energy Survey and other surveys; the provenance of each object and its half-light radius and luminosity estimates can be found in Table B1 of \cite{Kravtsov_2023}. 
We eliminate the Large Magellanic Cloud from our analysis with $(M_{\rm tot}(<r_{1/2}),L_V) = (2.8\times 10^{8} M_{\odot}, 1.5\times 10^{9} L_{V,\odot})$, because its total mass is estimated only approximately from its rotation curve, which is different from the mass estimates of all other galaxies. 
From these observations, we estimate the total mass within the half-light radius using the estimator of Ref.~\cite{Wolf_2010}
\begin{equation}
    \mathcal{M}_{\rm tot}(<r_{1/2})=930 \sigma^2_{\star,\text{los}}R_{1/2}M_{\odot},
\end{equation}
with $\sigma^2_{\star,\text{los}}$ the line of sight velocity dispersion of stars, $R_{1/2}$ the projected half-light radius, and $r_{1/2}$ the 3D stellar half-mass radius. 

\begin{figure}[ht!]
    \centering
    \includegraphics[width=1\textwidth]{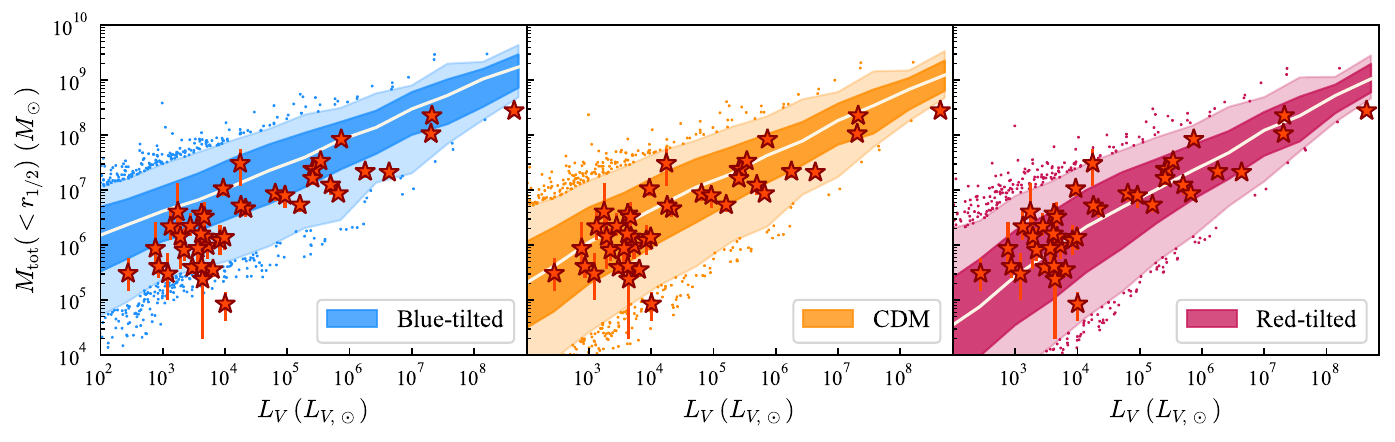}
  \caption{The total mass within half-light radius as a function of $V$-band galaxy luminosity for observed dwarf Milky Way satellites (red stars) and model galaxies (dots, line, and shaded bands). The mass in the model galaxies is computed assuming a blue-tilt (left panel), $\Lambda$CDM (middle panel) and red-tilt (right panel). The dots show the individual model galaxies, whereas the colored bands represent the 1 and 2 $\sigma$ regions of mass at a given $L_V$, and the solid white line shows the medians of the mass distribution. 
}\label{fig2}
\end{figure}

Figure~\ref{fig2} shows the observed Milky Way dwarf galaxy's total masses within half-light radii as a function of their V-band luminosity with red stars. The figure shows a clear correlation between $\mathcal{M}_{tot}(<r_{1/2})$ and $L_V$. The three panels compare model predictions for a blue-tilted spectrum with $(k_p,m_s)=(5,2)$ (left), $\Lambda$CDM (center), and a red-tilted spectrum with $(k_p,m_s)=(5,0)$ (right). The dots show the individual model galaxies, whereas the colored bands represent the 1 and 2 $\sigma$ regions of mass at a given $L_V$, and the solid white line shows the medians of the mass distribution. The figure shows that galaxies in the blue (red) tilted model have enclosed masses significantly larger (smaller) 
than the observed dwarf galaxies for fixed V-band luminosity, while the $\Lambda$CDM model describes the observed masses well.

In this study, we constrain models based on the observed and predicted $M_{\rm tot}(<r_{1/2})-L_V$ correlation and constrain the power law deviations from the $\Lambda$CDM spectrum using the statistical framework described below. 

\section{\label{sec3} Analysis}

We perform an unbinned log-likelihood ratio test. 
The log-likelihood for each cosmological model with parameters $\theta=(k_p,m_s)$ is constructed as follows,
\begin{align}
    \ln \mathcal{L} (\theta) = \sum_{i,\rm obs} \ln \left[
    \int \int \mathcal{P}(M_{\rm tot},L_V|,\theta) 
    \frac{1}{\sqrt{2\pi} \sigma_{M_{{\rm tot},i}}}
    \exp \left(-\frac{(M_{\rm tot}-M_{{\rm tot},i})^2}{2\sigma_{M_{{\rm tot},i}}^2}\right) \right.\\
    \times 
    \left.\frac{1}{\sqrt{2\pi} \sigma_{L_{V,i}}}
    \exp \left(-\frac{(L_{V}-L_{V,i})^2}{2\sigma_{L_{V,i}}^2}\right)
     dM_{\rm tot}dL_V  \right] ,
\end{align}
where $M_{tot}=M_{tot}(<r_{1/2})$. The two Gaussian distributions take into account observational uncertainties for each observed galaxy $i$. 
$\mathcal{P}(M_{\rm tot},L_V|,\theta)$ is the probability that a galaxy has true values ($M_{\rm tot},L_V$) under model parameter $\theta$, obtained using model galaxies shown in Fig.~\ref{fig2}. 

The continuous probability distribution of the model galaxies is estimated using a kernel density estimator (KDE). Specifically, for $x=(M_{\rm tot},L_V)$ the density of model galaxies is estimated  as
\begin{equation}
    \mathcal{P}(x)=\sum_{j,\rm sim} \frac{w_j}{h}\cdot K\left(\frac{x-X_j}{h} \right),
\end{equation}
where $K$ is the kernel function for which we adopt a Gaussian distribution with kernel bandwidth $h$, estimated using Scott's method~\cite{scott1992} as implemented in scikit-learn package~\cite{scikit-learn}, and $X_{j}=(M_{\rm tot,j},L_{V,j})$ for each modeled galaxy $j$.

In practice, probability of a given galaxy to be detected in a survey depends on their size, luminosity, and distance. 
To take into account detection probability of different model galaxies, we apply a weight $w_j$ to each modeled galaxy that reflects its typical detection probability, obtained by averaging over its different possible sky positions, as described in Section 2.6 of \cite{Manwadkar_2022}. 

We exclude model parameter $\theta$ that deviates significantly from $\Lambda$CDM following Wilk's theorem~\cite{Wilks:1938dza}, 
\begin{equation}
    \rm{TS} = -2\ln \left[ \frac{\mathcal{L}(\theta)}{\mathcal{L}(\Lambda\rm{CDM})} \right],
\end{equation}
where $\rm{TS} \geq  3.841$ is excluded at 95\% confidence level (CL). 

\section{\label{sec4} Results}

\begin{figure*}[ht!]
    \centering
    {{\includegraphics[width=0.45\textwidth]{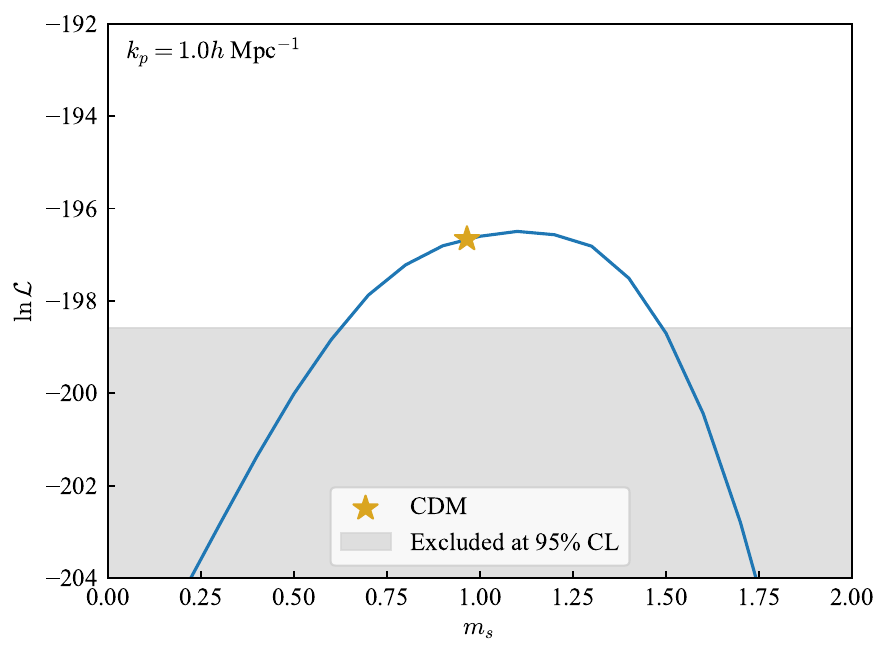} }}
    \quad
    {{\includegraphics[width=0.45\textwidth]{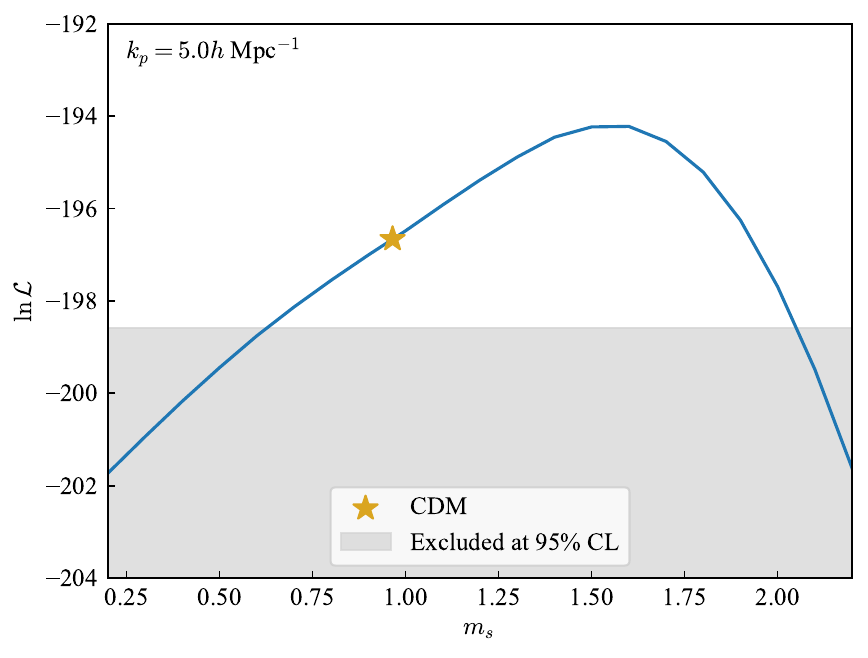} }}    
\caption{Log-likelihood as a function of the tilt $m_s$ (blue solid line) for $k_p=1.0h\rm\,Mpc^{-1}$ (left) and $k_p=5.0h\rm\,Mpc^{-1}$ (right). 
}\label{fig3}
\end{figure*}

The blue solid line in Fig.~\ref{fig3} shows the log-likelihood as a function of $m_s$ for pivot scales $k_p=1.0h\rm\,Mpc^{-1}$ (left) and $k_p=5.0h\rm\,Mpc^{-1}$ (right). The yellow star indicates $m_s=n_s$ ($\Lambda$CDM), and the grey area shows the region excluded at 95\% CL.  
Since we use the fiducial parameters of the galaxy formation model that were tested against observations assuming $\Lambda$CDM, we only consider significant deviations relative to $\Lambda$CDM. 
We leave it for future work to vary galaxy model parameters freely within allowed regions while remaining consistent with dwarf galaxy observations to obtain more robust constraints. 
This would also allow taking the maximum log-likelihood, which we expect to result in stronger constraints for both suppressed and enhanced models, as the data has a slight preference for a blue tilt for increasing pivot scale. 

\begin{figure*}[ht!]
    \centering
{{\includegraphics[width=0.55\textwidth]{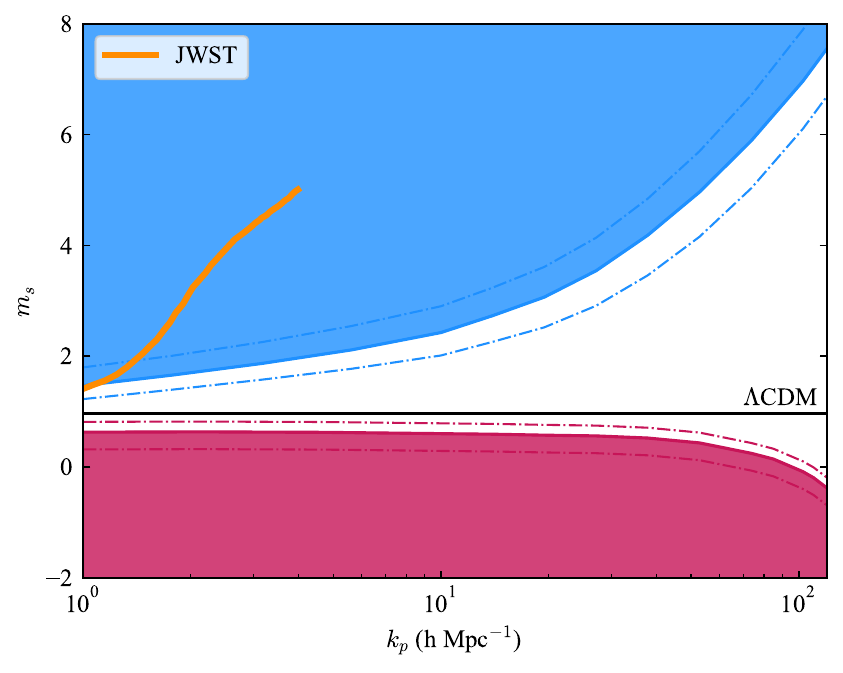} }}
\caption{Upper and lower 95\% CL limit on the tilt $m_s$ as a function of the pivot scale $k_p$. The dashed-dotted lines represent the error budget when considering uncertainties in the relations $r_{1/2}-M_{\star}$ and $M_{\star}-M_{200}$. The orange line shows the parameter space proposed to explain the JWST observations with a blue-tilt with typical star formation efficiency of $\epsilon=0.2$~\cite{Parashari_2023}. The grey horizontal band shows $m_s=n_s=0.9649\pm 0.0042$.
}
\label{fig4}
\end{figure*}

Figure~\ref{fig4} shows the 95\% excluded regions for the blue-tilted and red-tilted spectra on the amount of tilt $m_s$ as a function of the pivot scale $k_p$. Due to the skewed log-likelihood distribution at $k_p\gtrsim 3h\rm\, Mpc^{-1}$, the allowed region is asymmetric around $m_s=n_s$. The constraints can be used to explore implications for specific inflation models, which we defer for future work. 

We test uncertainties in the $r_{1/2}-M_{\star}$ and $M_{\star}-M_{200}$ relations that are predicted with the \texttt{GRUMPY} galaxy formation model. The former relation is constrained by observations, and Ref.~\cite{Kravtsov:2024esa} shows that indeed, the model used in this work reproduces the observed relation and its scatter. The modeled $r_{1/2}$ can vary by at most 30\% for a given $M_{\star}$, which we consider in the total error budget. 
Moreover, the $M_{\star}-M_{200}$ relation is constrained by the observed luminosity function of Milky Way satellite galaxies and is subject to approximately 30\% uncertainty due to Poisson noise and the halo-to-halo scatter in the model~\cite{Manwadkar_2022}. 
For each uncertainty, we obtain the excluded limit on $m_s$ at 95\% CL, and combine these by quadrature to obtain the total error budget at each pivot scale $k_p$, as represented in dashed-dotted lines in Fig.~\ref{fig4}. 

The orange line shows the parameter space that was proposed to explain the Cosmic Evolution Early Release Science Survey (CEERS) of the James Webb Space Telescope (JWST)~\cite{Bagley_2023} with a blue-tilt, obtained by~\cite{Parashari_2023}. Six galaxies at $7.4\leq z \leq 9.1$ were observed with stellar masses of $> 10^{10}M_{\odot}$~\cite{2023Natur.616..266L}. The constraint of \cite{Parashari_2023} was obtained assuming a constant star formation efficiency of $\epsilon=0.2$, defined as $M_{\star}=\epsilon \Omega_{\rm b}/\Omega_{\rm m}\,M_{\rm halo}$. 
Dwarf satellites analyzed in this study rule out the parameter space with pivot scales $k_p\gtrsim 1h\rm\,Mpc^{-1}$.
We note that the observed abundance of high-redshift galaxies can be reproduced within the $\Lambda$CDM model using reasonable choices for galaxy formation model \cite{chworowsky2023evidence,sun2023bursty}. 

\subsection{Linear matter power spectrum}
\begin{figure*}[ht!]
    \centering
{{\includegraphics[width=0.65\textwidth]{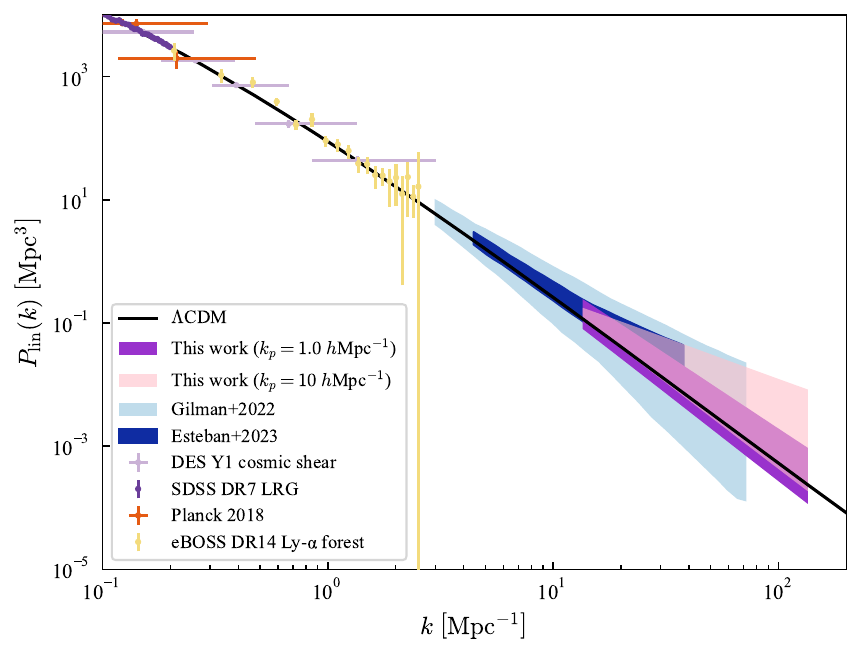} }}
\caption{Linear matter power spectrum at $z=0$. The 68\% contour band for a red- and blue-tilted spectrum obtained in this work are shown as the pink shaded regions for models with $k_p=1.0h\rm\,Mpc^{-1}$ and $k_p=10h\rm\,Mpc^{-1}$, compared with previous observations with DES, Planck, SDSS and eBOSS~\cite{Troxel_2018,2020,2017AJ....154...28B,Chabanier_2019}. Moreover, shown are the previous 68\% upper limits based on galaxy's velocity observations (dark blue)~\cite{esteban2023milky} and 68\% upper and lower limit based on strong gravitational lensing (light blue)~\cite{Gilman_2022}. }
\label{fig5}
\end{figure*}

Figure~\ref{fig5} shows the allowed linear matter power spectrum in shaded pink and purple at 68\% CL for pivot scales $k_p=1.0h\rm\,Mpc^{-1}$ and $10h\rm\,Mpc^{-1}$, which corresponds to $m_s=1.2$ (blue-tilt), $m_s=0.65$ (red-tilt) and $m_s=2.3$ (blue-tilt), $m_s=0.84$ (red-tilt), respectively. 
The faintest and brightest satellite galaxies in this analysis are Segue~I and the Small Magellanic Cloud (SMC) with $M_V=-1.30\pm 0.73$ and $M_V=-16.8\pm 0.1$, respectively~\cite{Kravtsov_2023}. Using the scaling relation $M_V$-$M_{h,\text{peak}}$ for the fiducial model from Ref.~\cite{2022MNRAS.514.2667K}, Segue~1 and SMC have peak halo mass of $M_{h,\text{peak}} \sim 10^8 M_{\odot}$ and $M_{h,\text{peak}} \sim 10^{11} M_{\odot}$. The corresponding wavenumber that we probe in our analysis can thus be estimated as $k_{\rm max}\sim 135$~Mpc$^{-1}$ and $k_{\rm min}\sim 13.5$~Mpc$^{-1}$ using Eq.~\ref{eq1}. This is illustrated by the shaded area in Fig.~\ref{fig5}. 

Figure~\ref{fig5} shows alongside our constraints previous analysis with DES~\cite{Troxel_2018}, Planck~\cite{2020}, SDSS~\cite{2017AJ....154...28B} and eBOSS Lyman-$\alpha$ forest~\cite{Chabanier_2019}, which are sensitive up to $k\lesssim 2.5$~Mpc$^{-1}$. Note that recent Lyman-$\alpha$ measurements of high-resolution spectra of quasars extend to much smaller scales of $k\sim 30$~Mpc$^{-1}$~\cite{Irsic:2023equ}. 
The light blue shaded area shows the allowed region using strong gravitational lensing measurements at 68\% CL~\cite{Gilman_2022}, which we improve on for both the red- and blue-tilted spectra. Moreover, the dark blue shaded area shows the allowed upper region at 68\% CL of Ref.~\cite{esteban2023milky} based on a joint analysis of internal velocities, sizes and total abundance of SDSS Milky Way satellites. We extend the analysis down to significant smaller scales.

\begin{figure*}[ht!]
    \centering
{{\includegraphics[width=0.65\textwidth]{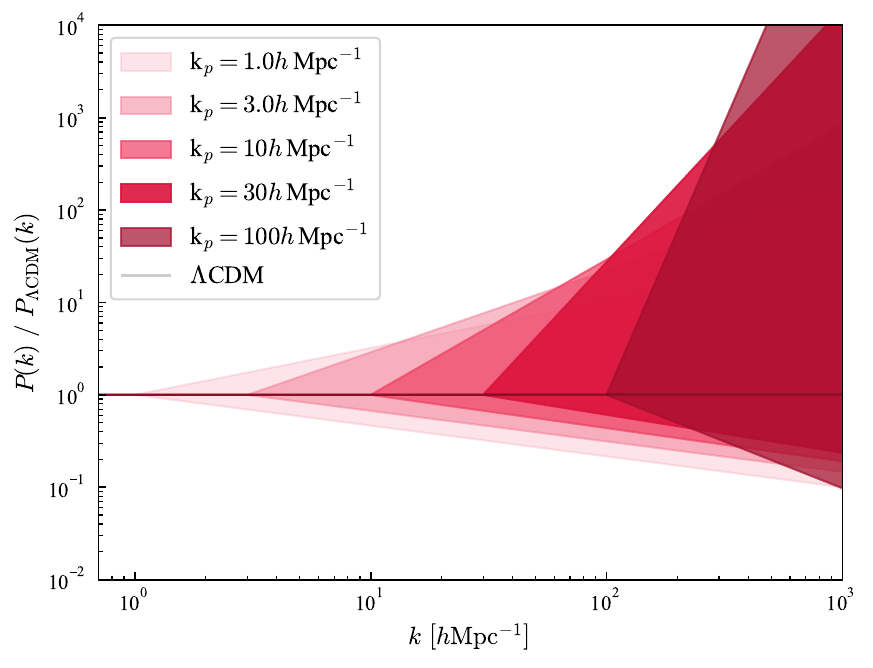} }}
\caption{The allowed 95\% CL of the matter power spectrum, normalized by the $\Lambda$CDM matter power spectrum, for different models of $k_p$. }
\label{fig6}
\end{figure*}

Figure~\ref{fig6} shows the amplitude of modeled linear power spectra that are allowed at 95\% CL, normalized by the $\Lambda$CDM linear power spectrum for different values of pivot scale $k_p$. The constraints are asymmetric around $\Lambda$CDM at larger pivot scales due to the slight preference for a blue-tilt. 

\subsection{Primordial power spectrum}

\begin{figure*}[ht!]
    \centering
{{\includegraphics[width=0.65\textwidth]{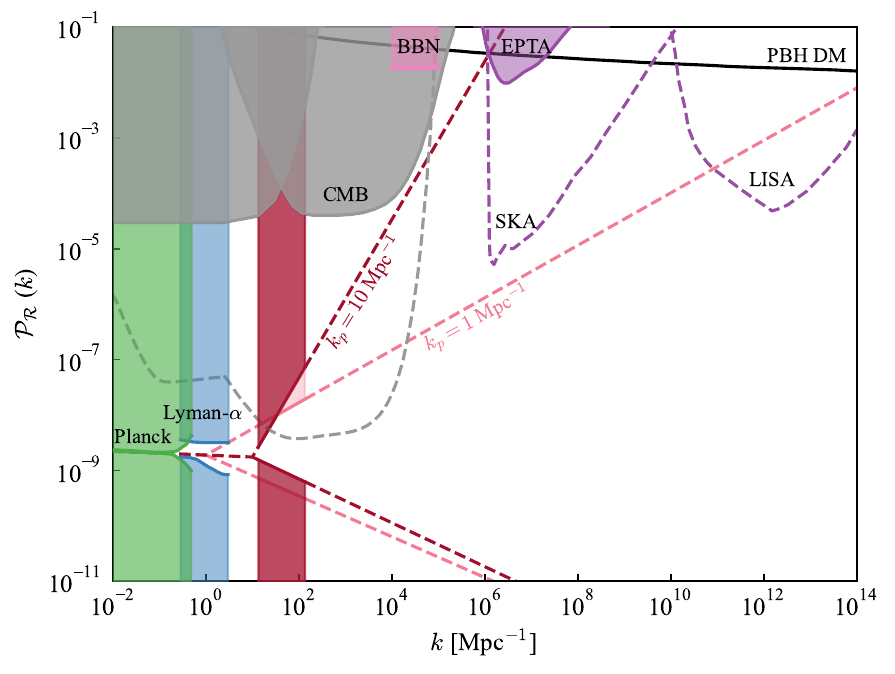} }}
\caption{Exclusion regions of the primordial power spectrum amplitude as a function of wavenumber $k$. 
}
\label{fig7}
\end{figure*}

Figure~\ref{fig7} shows the 95\% CL exclusion limit on the primordial curvature power spectrum obtained in this work as shaded pink areas for the models $k_p=1.0$~Mpc$^{-1}$ and $k_p=10$~Mpc$^{-1}$. The dashed lines extrapolates the model constrained in this work to scales that are beyond observational reach. 

Enhancements in primordial fluctuations can induce a stochastic gravitational wave background, which can be detected with gravitational wave detectors~\cite{Domenech:2021ztg,Ebadi:2023xhq}. 
Indeed, current observations with EPTA set limits as shown as the shaded purple region~\cite{EPTA:2015qep}, and already rule out blue-tilted models for large $k_p$. Future observations by SKA and LISA (dashed purple line) will significantly improve the current constraints~\cite{Inomata:2018epa,Byrnes:2018txb}. Moreover, recent results of pulsar timing arrays find evidence for a stochastic gravitational wave background at scales of $k \sim 10^6- 10^7$~Mpc$^{-1}$, which currently could be explained either by a blue-tilt or standard inspiraling supermassive black hole binaries~\cite{Afzal_2023}. 
If a blue-tilt explanation can be confirmed in the future, our analysis can provide insights in what inflationary models are viable, as it narrows down the allowed amplitudes at different wavelengths. 

PBHs are typically predicted to be formed through collapse of overdensities during radiation domination~\cite{Green:2020jor}. An excess in the amplitude of the primordial power spectrum therefore leads to an enhancement of PBH production. 
For PBH to make up all of dark matter, large fluctuations are required, as is shown as the black line in Fig.~\ref{fig7} from Ref.~\cite{Carr:2020gox}. 
The PBH mass that corresponds to the wavelengths that we're sensitive to ($M_{\text{PBH}}\sim 10^8-10^{12}$M$_{\odot}$~\cite{Kawasaki:2016pql}) is however already ruled out as it would modify the recombination history of the Universe~\cite{Green:2020jor}, and current observations show that PBH could only consist out of a tiny fraction of dark matter.  
Interestingly, extrapolating the blue-tilt model to small scales leads to parameter spaces in which PBH are viable as dark matter, which might be testable when combining with future observations. 

Figure~\ref{fig7} shows alongside constraints from CMB spectral distortions in grey from COBE/FIRAS measurements (shaded)~\cite{Chluba:2012we,Mather:1993ij,Fixsen:1996nj} and future sensitivity with a PIXIE-type satellite (dashed)~\cite{Cyr:2023pgw}. The green shaded area comes from analyzing the CMB temperature angular power spectrum~\cite{Planck:2018jri} and the blue shaded area from the Lyman-$\alpha$ forest~\cite{Bird_2011}. 
Moreover, the pink constraints are derived from the impact of large density perturbations on the abundances of light elements, modifying Big Bang nucleosynthesis~\cite{Inomata:2016uip}. 

Finally, Ref.~\cite{Graham:2024hah} set strong constraints on the primordial power spectrum in the $k$--range of $\sim 10-1000$~Mpc$^{-1}$ by considering the dynamical heating effect dark matter substructure in the $\sim 10-10^8 M_{\odot}$ mass-scale has on stars in ultra-faint dwarf galaxies. Previous work has however shown that only the most massive substructure plays a dynamical role in the heating of stars in the Galactic disk~\cite{Font:2001py}. Thus, the constraints from dynamical heating cannot extend
to much smaller scales than the scales probed by host subhalos themselves. 

\section{Conclusions}
We use measured velocity dispersions, half-light radii and $V$-band luminosities of 41 observed dwarf galaxies to set stringent limits on the amplitude and slope of the linear and primordial power spectrum at scales $k\approx 1-100$~Mpc$^{-1}$. 
Deviations in the primordial fluctuations from the standard $\Lambda$CDM model are expected from various non-standard inflationary and dark matter models. Such deviations should lead to different halo concentrations, and, consequently, affect the central mass of dwarf galaxies.  The wavelengths corresponding to the halo concentrations of the faintest and brightest galaxies in our sample are $k\sim 13.5$~Mpc$^{-1}$ and $k\sim 135$~Mpc$^{-1}$, respectively. Therefore, this analysis serves as a powerful tool to test the features in the power spectrum at small scales, where only a few studies have explored both enhancement and suppression of primordial fluctuations. 

In this work, we model galaxies by combining high-resolution numerical simulations with a realistic galaxy formation model. We used the fiducial parameters of the model that have been validated against various observations of Milky-Way satellite galaxies assuming $\Lambda$CDM. We show that the estimates of the central densities within half-light radii of observed dwarf galaxies constrain the power spectrum to be close to the $\Lambda$CDM prediction on scales probed by the galaxies. 

By using fixed galaxy formation model parameters, the analysis is limited to only rule out models that significant deviate with respect to $\Lambda$CDM based on differences in concentration of dark matter halos. Marginalizing over galaxy formation parameter might improve the analysis, however this requires validation against a wide range of observational data. 

Blue- and red-tilted cosmologies might affect other galaxy properties beyond their central densities. For instance, an earlier onset of structure formation in blue-tilted models could lead to a larger population of halos that can host galaxies prior to reionization, beyond which the formation of faint galaxies is suppressed. This would predict an increase in number of faint satellite galaxies, as well as brighter dwarf galaxies. Testing this would require running numerical simulations for each cosmology, coupled with galaxy formation modeling. 
This approach is computational expensive, and has not been well studied for blue tilted models due to computational challenges resolving the enhanced small scale power. Further exploration could improve the sensitivity in testing cosmological models, however it is beyond the scope of the current work and is left for future work.

The continuous tilt in the power spectrum tested in this work can be extended to specific models that predict peaks or bumps in the range of $10\text{ Mpc}^{-1}\lesssim k\lesssim 100\text{ Mpc}^{-1}$. Additionally, models with a continuous tilt can be further assessed by combining our results with measurements at different wavelengths, such as those from pulsar timing arrays. These complementary searches will provide a deeper understanding of viable models and are crucial to differentiate between astrophysical or non-standard $\Lambda$CDM scenarios. Furthermore, upcoming kinematic data from the Dark Energy Survey and Rubin Observatory observations will become available for a much larger number of low-redshift galaxies which will advance our understanding even further. 

\acknowledgments{We thank the anonymous referee for their detailed feedback that improved this work. We are grateful to Akhil Premkumar for helpful discussions. AD is supported by the Kavli Institute for Cosmological physics at the University of Chicago through an endowment from the Kavli Foundation and its founder Fred Kavli. }

\bibliography{References}

\begin{thebibliography}{96}%
\makeatletter
\providecommand \@ifxundefined [1]{%
 \@ifx{#1\undefined}
}%
\providecommand \@ifnum [1]{%
 \ifnum #1\expandafter \@firstoftwo
 \else \expandafter \@secondoftwo
 \fi
}%
\providecommand \@ifx [1]{%
 \ifx #1\expandafter \@firstoftwo
 \else \expandafter \@secondoftwo
 \fi
}%
\providecommand \natexlab [1]{#1}%
\providecommand \enquote  [1]{``#1''}%
\providecommand \bibnamefont  [1]{#1}%
\providecommand \bibfnamefont [1]{#1}%
\providecommand \citenamefont [1]{#1}%
\providecommand \href@noop [0]{\@secondoftwo}%
\providecommand \href [0]{\begingroup \@sanitize@url \@href}%
\providecommand \@href[1]{\@@startlink{#1}\@@href}%
\providecommand \@@href[1]{\endgroup#1\@@endlink}%
\providecommand \@sanitize@url [0]{\catcode `\\12\catcode `\$12\catcode `\&12\catcode `\#12\catcode `\^12\catcode `\_12\catcode `\%12\relax}%
\providecommand \@@startlink[1]{}%
\providecommand \@@endlink[0]{}%
\providecommand \url  [0]{\begingroup\@sanitize@url \@url }%
\providecommand \@url [1]{\endgroup\@href {#1}{\urlprefix }}%
\providecommand \urlprefix  [0]{URL }%
\providecommand \Eprint [0]{\href }%
\providecommand \doibase [0]{https://doi.org/}%
\providecommand \selectlanguage [0]{\@gobble}%
\providecommand \bibinfo  [0]{\@secondoftwo}%
\providecommand \bibfield  [0]{\@secondoftwo}%
\providecommand \translation [1]{[#1]}%
\providecommand \BibitemOpen [0]{}%
\providecommand \bibitemStop [0]{}%
\providecommand \bibitemNoStop [0]{.\EOS\space}%
\providecommand \EOS [0]{\spacefactor3000\relax}%
\providecommand \BibitemShut  [1]{\csname bibitem#1\endcsname}%
\let\auto@bib@innerbib\@empty
\bibitem [{\citenamefont {{Mukhanov}}\ and\ \citenamefont {{Chibisov}}(1981)}]{Mukhanov.Chibisov.1981}%
  \BibitemOpen
  \bibfield  {author} {\bibinfo {author} {\bibfnamefont {V.~F.}\ \bibnamefont {{Mukhanov}}}\ and\ \bibinfo {author} {\bibfnamefont {G.~V.}\ \bibnamefont {{Chibisov}}},\ }\bibfield  {title} {\bibinfo {title} {{Quantum fluctuations and a nonsingular universe}},\ }\href@noop {} {\bibfield  {journal} {\bibinfo  {journal} {Soviet Journal of Experimental and Theoretical Physics Letters}\ }\textbf {\bibinfo {volume} {33}},\ \bibinfo {pages} {532} (\bibinfo {year} {1981})}\BibitemShut {NoStop}%
\bibitem [{\citenamefont {{Starobinsky}}(1982)}]{Starobinsky.1982}%
  \BibitemOpen
  \bibfield  {author} {\bibinfo {author} {\bibfnamefont {A.~A.}\ \bibnamefont {{Starobinsky}}},\ }\bibfield  {title} {\bibinfo {title} {{Dynamics of phase transition in the new inflationary universe scenario and generation of perturbations}},\ }\href {https://doi.org/10.1016/0370-2693(82)90541-X} {\bibfield  {journal} {\bibinfo  {journal} {Physics Letters B}\ }\textbf {\bibinfo {volume} {117}},\ \bibinfo {pages} {175} (\bibinfo {year} {1982})}\BibitemShut {NoStop}%
\bibitem [{\citenamefont {{Linde}}(1982)}]{Linde.1982}%
  \BibitemOpen
  \bibfield  {author} {\bibinfo {author} {\bibfnamefont {A.~D.}\ \bibnamefont {{Linde}}},\ }\bibfield  {title} {\bibinfo {title} {{Scalar field fluctuations in the expanding universe and the new inflationary universe scenario}},\ }\href {https://doi.org/10.1016/0370-2693(82)90293-3} {\bibfield  {journal} {\bibinfo  {journal} {Physics Letters B}\ }\textbf {\bibinfo {volume} {116}},\ \bibinfo {pages} {335} (\bibinfo {year} {1982})}\BibitemShut {NoStop}%
\bibitem [{\citenamefont {{Hawking}}(1982)}]{Hawking.1982}%
  \BibitemOpen
  \bibfield  {author} {\bibinfo {author} {\bibfnamefont {S.~W.}\ \bibnamefont {{Hawking}}},\ }\bibfield  {title} {\bibinfo {title} {{The development of irregularities in a single bubble inflationary universe}},\ }\href {https://doi.org/10.1016/0370-2693(82)90373-2} {\bibfield  {journal} {\bibinfo  {journal} {Physics Letters B}\ }\textbf {\bibinfo {volume} {115}},\ \bibinfo {pages} {295} (\bibinfo {year} {1982})}\BibitemShut {NoStop}%
\bibitem [{\citenamefont {Baumann}(2012)}]{baumann2012tasi}%
  \BibitemOpen
  \bibfield  {author} {\bibinfo {author} {\bibfnamefont {D.}~\bibnamefont {Baumann}},\ }\href@noop {} {\bibinfo {title} {Tasi lectures on inflation}} (\bibinfo {year} {2012}),\ \Eprint {https://arxiv.org/abs/0907.5424} {arXiv:0907.5424 [hep-th]} \BibitemShut {NoStop}%
\bibitem [{\citenamefont {Aghanim}\ \emph {et~al.}(2020)\citenamefont {Aghanim} \emph {et~al.}}]{2020}%
  \BibitemOpen
  \bibfield  {author} {\bibinfo {author} {\bibfnamefont {N.}~\bibnamefont {Aghanim}} \emph {et~al.} (\bibinfo {collaboration} {Planck}),\ }\bibfield  {title} {\bibinfo {title} {{Planck 2018 results. VI. Cosmological parameters}},\ }\href {https://doi.org/10.1051/0004-6361/201833910} {\bibfield  {journal} {\bibinfo  {journal} {Astron. Astrophys.}\ }\textbf {\bibinfo {volume} {641}},\ \bibinfo {pages} {A6} (\bibinfo {year} {2020})},\ \bibinfo {note} {[Erratum: Astron.Astrophys. 652, C4 (2021)]},\ \Eprint {https://arxiv.org/abs/1807.06209} {arXiv:1807.06209 [astro-ph.CO]} \BibitemShut {NoStop}%
\bibitem [{\citenamefont {Troxel}\ \emph {et~al.}(2018)\citenamefont {Troxel} \emph {et~al.}}]{Troxel_2018}%
  \BibitemOpen
  \bibfield  {author} {\bibinfo {author} {\bibfnamefont {M.~A.}\ \bibnamefont {Troxel}} \emph {et~al.} (\bibinfo {collaboration} {DES}),\ }\bibfield  {title} {\bibinfo {title} {{Dark Energy Survey Year 1 results: Cosmological constraints from cosmic shear}},\ }\href {https://doi.org/10.1103/PhysRevD.98.043528} {\bibfield  {journal} {\bibinfo  {journal} {Phys. Rev. D}\ }\textbf {\bibinfo {volume} {98}},\ \bibinfo {pages} {043528} (\bibinfo {year} {2018})},\ \Eprint {https://arxiv.org/abs/1708.01538} {arXiv:1708.01538 [astro-ph.CO]} \BibitemShut {NoStop}%
\bibitem [{\citenamefont {Blanton}\ \emph {et~al.}(2017)\citenamefont {Blanton} \emph {et~al.}}]{2017AJ....154...28B}%
  \BibitemOpen
  \bibfield  {author} {\bibinfo {author} {\bibfnamefont {M.~R.}\ \bibnamefont {Blanton}} \emph {et~al.} (\bibinfo {collaboration} {eBOSS}),\ }\bibfield  {title} {\bibinfo {title} {{Sloan Digital Sky Survey IV: Mapping the Milky Way, Nearby Galaxies and the Distant Universe}},\ }\href {https://doi.org/10.3847/1538-3881/aa7567} {\bibfield  {journal} {\bibinfo  {journal} {Astron. J.}\ }\textbf {\bibinfo {volume} {154}},\ \bibinfo {pages} {28} (\bibinfo {year} {2017})},\ \Eprint {https://arxiv.org/abs/1703.00052} {arXiv:1703.00052 [astro-ph.GA]} \BibitemShut {NoStop}%
\bibitem [{\citenamefont {Chabanier}\ \emph {et~al.}(2019)\citenamefont {Chabanier}, \citenamefont {Millea},\ and\ \citenamefont {Palanque-Delabrouille}}]{Chabanier_2019}%
  \BibitemOpen
  \bibfield  {author} {\bibinfo {author} {\bibfnamefont {S.}~\bibnamefont {Chabanier}}, \bibinfo {author} {\bibfnamefont {M.}~\bibnamefont {Millea}},\ and\ \bibinfo {author} {\bibfnamefont {N.}~\bibnamefont {Palanque-Delabrouille}},\ }\bibfield  {title} {\bibinfo {title} {Matter power spectrum: from ly $\alpha$ forest to cmb scales},\ }\href {https://doi.org/10.1093/mnras/stz2310} {\bibfield  {journal} {\bibinfo  {journal} {Monthly Notices of the Royal Astronomical Society}\ }\textbf {\bibinfo {volume} {489}},\ \bibinfo {pages} {2247–2253} (\bibinfo {year} {2019})}\BibitemShut {NoStop}%
\bibitem [{\citenamefont {Bartolo}\ \emph {et~al.}(2004)\citenamefont {Bartolo}, \citenamefont {Komatsu}, \citenamefont {Matarrese},\ and\ \citenamefont {Riotto}}]{Bartolo:2004if}%
  \BibitemOpen
  \bibfield  {author} {\bibinfo {author} {\bibfnamefont {N.}~\bibnamefont {Bartolo}}, \bibinfo {author} {\bibfnamefont {E.}~\bibnamefont {Komatsu}}, \bibinfo {author} {\bibfnamefont {S.}~\bibnamefont {Matarrese}},\ and\ \bibinfo {author} {\bibfnamefont {A.}~\bibnamefont {Riotto}},\ }\bibfield  {title} {\bibinfo {title} {{Non-Gaussianity from inflation: Theory and observations}},\ }\href {https://doi.org/10.1016/j.physrep.2004.08.022} {\bibfield  {journal} {\bibinfo  {journal} {Phys. Rept.}\ }\textbf {\bibinfo {volume} {402}},\ \bibinfo {pages} {103} (\bibinfo {year} {2004})},\ \Eprint {https://arxiv.org/abs/astro-ph/0406398} {arXiv:astro-ph/0406398} \BibitemShut {NoStop}%
\bibitem [{\citenamefont {Bridle}\ \emph {et~al.}(2003)\citenamefont {Bridle}, \citenamefont {Lewis}, \citenamefont {Weller},\ and\ \citenamefont {Efstathiou}}]{Bridle_2003}%
  \BibitemOpen
  \bibfield  {author} {\bibinfo {author} {\bibfnamefont {S.~L.}\ \bibnamefont {Bridle}}, \bibinfo {author} {\bibfnamefont {A.~M.}\ \bibnamefont {Lewis}}, \bibinfo {author} {\bibfnamefont {J.}~\bibnamefont {Weller}},\ and\ \bibinfo {author} {\bibfnamefont {G.}~\bibnamefont {Efstathiou}},\ }\bibfield  {title} {\bibinfo {title} {Reconstructing the primordial power spectrum},\ }\href {https://doi.org/10.1046/j.1365-8711.2003.06807.x} {\bibfield  {journal} {\bibinfo  {journal} {Monthly Notices of the Royal Astronomical Society}\ }\textbf {\bibinfo {volume} {342}},\ \bibinfo {pages} {L72–L78} (\bibinfo {year} {2003})}\BibitemShut {NoStop}%
\bibitem [{\citenamefont {Bechtol}\ \emph {et~al.}(2022)\citenamefont {Bechtol} \emph {et~al.}}]{Bechtol:2022koa}%
  \BibitemOpen
  \bibfield  {author} {\bibinfo {author} {\bibfnamefont {K.}~\bibnamefont {Bechtol}} \emph {et~al.},\ }\bibfield  {title} {\bibinfo {title} {{Snowmass2021 Cosmic Frontier White Paper: Dark Matter Physics from Halo Measurements}},\ }in\ \href@noop {} {\emph {\bibinfo {booktitle} {{Snowmass 2021}}}}\ (\bibinfo {year} {2022})\ \Eprint {https://arxiv.org/abs/2203.07354} {arXiv:2203.07354 [hep-ph]} \BibitemShut {NoStop}%
\bibitem [{\citenamefont {Tulin}\ and\ \citenamefont {Yu}(2018)}]{Tulin:2017ara}%
  \BibitemOpen
  \bibfield  {author} {\bibinfo {author} {\bibfnamefont {S.}~\bibnamefont {Tulin}}\ and\ \bibinfo {author} {\bibfnamefont {H.-B.}\ \bibnamefont {Yu}},\ }\bibfield  {title} {\bibinfo {title} {{Dark Matter Self-interactions and Small Scale Structure}},\ }\href {https://doi.org/10.1016/j.physrep.2017.11.004} {\bibfield  {journal} {\bibinfo  {journal} {Phys. Rept.}\ }\textbf {\bibinfo {volume} {730}},\ \bibinfo {pages} {1} (\bibinfo {year} {2018})},\ \Eprint {https://arxiv.org/abs/1705.02358} {arXiv:1705.02358 [hep-ph]} \BibitemShut {NoStop}%
\bibitem [{\citenamefont {Padilla}\ \emph {et~al.}(2019)\citenamefont {Padilla}, \citenamefont {Vázquez}, \citenamefont {Matos},\ and\ \citenamefont {Germán}}]{Padilla_2019}%
  \BibitemOpen
  \bibfield  {author} {\bibinfo {author} {\bibfnamefont {L.~E.}\ \bibnamefont {Padilla}}, \bibinfo {author} {\bibfnamefont {J.~A.}\ \bibnamefont {Vázquez}}, \bibinfo {author} {\bibfnamefont {T.}~\bibnamefont {Matos}},\ and\ \bibinfo {author} {\bibfnamefont {G.}~\bibnamefont {Germán}},\ }\bibfield  {title} {\bibinfo {title} {Scalar field dark matter spectator during inflation: the effect of self-interaction},\ }\href {https://doi.org/10.1088/1475-7516/2019/05/056} {\bibfield  {journal} {\bibinfo  {journal} {Journal of Cosmology and Astroparticle Physics}\ }\textbf {\bibinfo {volume} {2019}}\bibinfo  {number} { (05)},\ \bibinfo {pages} {056–056}}\BibitemShut {NoStop}%
\bibitem [{\citenamefont {Biagetti}\ \emph {et~al.}(2013)\citenamefont {Biagetti}, \citenamefont {Fasiello},\ and\ \citenamefont {Riotto}}]{PhysRevD.88.103518}%
  \BibitemOpen
\bibfield  {number} {  }\bibfield  {author} {\bibinfo {author} {\bibfnamefont {M.}~\bibnamefont {Biagetti}}, \bibinfo {author} {\bibfnamefont {M.}~\bibnamefont {Fasiello}},\ and\ \bibinfo {author} {\bibfnamefont {A.}~\bibnamefont {Riotto}},\ }\bibfield  {title} {\bibinfo {title} {Enhancing inflationary tensor modes through spectator fields},\ }\href {https://doi.org/10.1103/PhysRevD.88.103518} {\bibfield  {journal} {\bibinfo  {journal} {Phys. Rev. D}\ }\textbf {\bibinfo {volume} {88}},\ \bibinfo {pages} {103518} (\bibinfo {year} {2013})}\BibitemShut {NoStop}%
\bibitem [{\citenamefont {Biagetti}\ \emph {et~al.}(2015)\citenamefont {Biagetti}, \citenamefont {Dimastrogiovanni}, \citenamefont {Fasiello},\ and\ \citenamefont {Peloso}}]{Biagetti_2015}%
  \BibitemOpen
  \bibfield  {author} {\bibinfo {author} {\bibfnamefont {M.}~\bibnamefont {Biagetti}}, \bibinfo {author} {\bibfnamefont {E.}~\bibnamefont {Dimastrogiovanni}}, \bibinfo {author} {\bibfnamefont {M.}~\bibnamefont {Fasiello}},\ and\ \bibinfo {author} {\bibfnamefont {M.}~\bibnamefont {Peloso}},\ }\bibfield  {title} {\bibinfo {title} {Gravitational waves and scalar perturbations from spectator fields},\ }\href {https://doi.org/10.1088/1475-7516/2015/04/011} {\bibfield  {journal} {\bibinfo  {journal} {Journal of Cosmology and Astroparticle Physics}\ }\textbf {\bibinfo {volume} {2015}}\bibinfo  {number} { (04)},\ \bibinfo {pages} {011–011}}\BibitemShut {NoStop}%
\bibitem [{\citenamefont {Iarygina}\ \emph {et~al.}(2020)\citenamefont {Iarygina}, \citenamefont {Sfakianakis}, \citenamefont {Wang},\ and\ \citenamefont {Ach\'ucarro}}]{Iarygina:2020dwe}%
  \BibitemOpen
\bibfield  {number} {  }\bibfield  {author} {\bibinfo {author} {\bibfnamefont {O.}~\bibnamefont {Iarygina}}, \bibinfo {author} {\bibfnamefont {E.~I.}\ \bibnamefont {Sfakianakis}}, \bibinfo {author} {\bibfnamefont {D.-G.}\ \bibnamefont {Wang}},\ and\ \bibinfo {author} {\bibfnamefont {A.}~\bibnamefont {Ach\'ucarro}},\ }\bibfield  {title} {\bibinfo {title} {{Multi-field inflation and preheating in asymmetric $\alpha$-attractors}},\ }\href@noop {} {\  (\bibinfo {year} {2020})},\ \Eprint {https://arxiv.org/abs/2005.00528} {arXiv:2005.00528 [astro-ph.CO]} \BibitemShut {NoStop}%
\bibitem [{\citenamefont {Wands}()}]{Wands}%
  \BibitemOpen
  \bibfield  {author} {\bibinfo {author} {\bibfnamefont {D.}~\bibnamefont {Wands}},\ }\bibinfo {title} {Multiple field inflation},\ in\ \href {https://doi.org/10.1007/978-3-540-74353-8_8} {\emph {\bibinfo {booktitle} {Lecture Notes in Physics}}}\ (\bibinfo  {publisher} {Springer Berlin Heidelberg})\ p.\ \bibinfo {pages} {275–304}\BibitemShut {NoStop}%
\bibitem [{\citenamefont {Ebadi}\ \emph {et~al.}(2023)\citenamefont {Ebadi}, \citenamefont {Kumar}, \citenamefont {McCune}, \citenamefont {Tai},\ and\ \citenamefont {Wang}}]{Ebadi:2023xhq}%
  \BibitemOpen
  \bibfield  {author} {\bibinfo {author} {\bibfnamefont {R.}~\bibnamefont {Ebadi}}, \bibinfo {author} {\bibfnamefont {S.}~\bibnamefont {Kumar}}, \bibinfo {author} {\bibfnamefont {A.}~\bibnamefont {McCune}}, \bibinfo {author} {\bibfnamefont {H.}~\bibnamefont {Tai}},\ and\ \bibinfo {author} {\bibfnamefont {L.-T.}\ \bibnamefont {Wang}},\ }\bibfield  {title} {\bibinfo {title} {{Gravitational Waves from Stochastic Scalar Fluctuations}},\ }\href@noop {} {\  (\bibinfo {year} {2023})},\ \Eprint {https://arxiv.org/abs/2307.01248} {arXiv:2307.01248 [astro-ph.CO]} \BibitemShut {NoStop}%
\bibitem [{\citenamefont {Anber}\ and\ \citenamefont {Sorbo}(2010)}]{PhysRevD.81.043534}%
  \BibitemOpen
  \bibfield  {author} {\bibinfo {author} {\bibfnamefont {M.~M.}\ \bibnamefont {Anber}}\ and\ \bibinfo {author} {\bibfnamefont {L.}~\bibnamefont {Sorbo}},\ }\bibfield  {title} {\bibinfo {title} {Naturally inflating on steep potentials through electromagnetic dissipation},\ }\href {https://doi.org/10.1103/PhysRevD.81.043534} {\bibfield  {journal} {\bibinfo  {journal} {Phys. Rev. D}\ }\textbf {\bibinfo {volume} {81}},\ \bibinfo {pages} {043534} (\bibinfo {year} {2010})}\BibitemShut {NoStop}%
\bibitem [{\citenamefont {Anber}\ and\ \citenamefont {Sorbo}(2012)}]{PhysRevD.85.123537}%
  \BibitemOpen
  \bibfield  {author} {\bibinfo {author} {\bibfnamefont {M.~M.}\ \bibnamefont {Anber}}\ and\ \bibinfo {author} {\bibfnamefont {L.}~\bibnamefont {Sorbo}},\ }\bibfield  {title} {\bibinfo {title} {Non-gaussianities and chiral gravitational waves in natural steep inflation},\ }\href {https://doi.org/10.1103/PhysRevD.85.123537} {\bibfield  {journal} {\bibinfo  {journal} {Phys. Rev. D}\ }\textbf {\bibinfo {volume} {85}},\ \bibinfo {pages} {123537} (\bibinfo {year} {2012})}\BibitemShut {NoStop}%
\bibitem [{\citenamefont {Chung}\ and\ \citenamefont {Yoo}(2015)}]{Chung_2015}%
  \BibitemOpen
  \bibfield  {author} {\bibinfo {author} {\bibfnamefont {D.~J.}\ \bibnamefont {Chung}}\ and\ \bibinfo {author} {\bibfnamefont {H.}~\bibnamefont {Yoo}},\ }\bibfield  {title} {\bibinfo {title} {Elementary theorems regarding blue isocurvature perturbations},\ }\bibfield  {journal} {\bibinfo  {journal} {Physical Review D}\ }\textbf {\bibinfo {volume} {91}},\ \href {https://doi.org/10.1103/physrevd.91.083530} {10.1103/physrevd.91.083530} (\bibinfo {year} {2015})\BibitemShut {NoStop}%
\bibitem [{\citenamefont {Talebian}\ \emph {et~al.}(2022)\citenamefont {Talebian}, \citenamefont {Nassiri-Rad},\ and\ \citenamefont {Firouzjahi}}]{Talebian:2022jkb}%
  \BibitemOpen
  \bibfield  {author} {\bibinfo {author} {\bibfnamefont {A.}~\bibnamefont {Talebian}}, \bibinfo {author} {\bibfnamefont {A.}~\bibnamefont {Nassiri-Rad}},\ and\ \bibinfo {author} {\bibfnamefont {H.}~\bibnamefont {Firouzjahi}},\ }\bibfield  {title} {\bibinfo {title} {{Stochastic effects in axion inflation and primordial black hole formation}},\ }\href {https://doi.org/10.1103/PhysRevD.105.103516} {\bibfield  {journal} {\bibinfo  {journal} {Phys. Rev. D}\ }\textbf {\bibinfo {volume} {105}},\ \bibinfo {pages} {103516} (\bibinfo {year} {2022})},\ \Eprint {https://arxiv.org/abs/2202.02062} {arXiv:2202.02062 [astro-ph.CO]} \BibitemShut {NoStop}%
\bibitem [{\citenamefont {Graham}\ \emph {et~al.}(2016)\citenamefont {Graham}, \citenamefont {Mardon},\ and\ \citenamefont {Rajendran}}]{Graham_2016}%
  \BibitemOpen
  \bibfield  {author} {\bibinfo {author} {\bibfnamefont {P.~W.}\ \bibnamefont {Graham}}, \bibinfo {author} {\bibfnamefont {J.}~\bibnamefont {Mardon}},\ and\ \bibinfo {author} {\bibfnamefont {S.}~\bibnamefont {Rajendran}},\ }\bibfield  {title} {\bibinfo {title} {Vector dark matter from inflationary fluctuations},\ }\bibfield  {journal} {\bibinfo  {journal} {Physical Review D}\ }\textbf {\bibinfo {volume} {93}},\ \href {https://doi.org/10.1103/physrevd.93.103520} {10.1103/physrevd.93.103520} (\bibinfo {year} {2016})\BibitemShut {NoStop}%
\bibitem [{\citenamefont {Barnaby}\ \emph {et~al.}(2012)\citenamefont {Barnaby}, \citenamefont {Moxon}, \citenamefont {Namba}, \citenamefont {Peloso}, \citenamefont {Shiu},\ and\ \citenamefont {Zhou}}]{PhysRevD.86.103508}%
  \BibitemOpen
  \bibfield  {author} {\bibinfo {author} {\bibfnamefont {N.}~\bibnamefont {Barnaby}}, \bibinfo {author} {\bibfnamefont {J.}~\bibnamefont {Moxon}}, \bibinfo {author} {\bibfnamefont {R.}~\bibnamefont {Namba}}, \bibinfo {author} {\bibfnamefont {M.}~\bibnamefont {Peloso}}, \bibinfo {author} {\bibfnamefont {G.}~\bibnamefont {Shiu}},\ and\ \bibinfo {author} {\bibfnamefont {P.}~\bibnamefont {Zhou}},\ }\bibfield  {title} {\bibinfo {title} {Gravity waves and non-gaussian features from particle production in a sector gravitationally coupled to the inflaton},\ }\href {https://doi.org/10.1103/PhysRevD.86.103508} {\bibfield  {journal} {\bibinfo  {journal} {Phys. Rev. D}\ }\textbf {\bibinfo {volume} {86}},\ \bibinfo {pages} {103508} (\bibinfo {year} {2012})}\BibitemShut {NoStop}%
\bibitem [{\citenamefont {Bartolo}\ \emph {et~al.}(2016)\citenamefont {Bartolo}, \citenamefont {Caprini}, \citenamefont {Domcke}, \citenamefont {Figueroa}, \citenamefont {Garcia-Bellido}, \citenamefont {Guzzetti}, \citenamefont {Liguori}, \citenamefont {Matarrese}, \citenamefont {Peloso}, \citenamefont {Petiteau}, \citenamefont {Ricciardone}, \citenamefont {Sakellariadou}, \citenamefont {Sorbo},\ and\ \citenamefont {Tasinato}}]{Bartolo_2016}%
  \BibitemOpen
  \bibfield  {author} {\bibinfo {author} {\bibfnamefont {N.}~\bibnamefont {Bartolo}}, \bibinfo {author} {\bibfnamefont {C.}~\bibnamefont {Caprini}}, \bibinfo {author} {\bibfnamefont {V.}~\bibnamefont {Domcke}}, \bibinfo {author} {\bibfnamefont {D.~G.}\ \bibnamefont {Figueroa}}, \bibinfo {author} {\bibfnamefont {J.}~\bibnamefont {Garcia-Bellido}}, \bibinfo {author} {\bibfnamefont {M.~C.}\ \bibnamefont {Guzzetti}}, \bibinfo {author} {\bibfnamefont {M.}~\bibnamefont {Liguori}}, \bibinfo {author} {\bibfnamefont {S.}~\bibnamefont {Matarrese}}, \bibinfo {author} {\bibfnamefont {M.}~\bibnamefont {Peloso}}, \bibinfo {author} {\bibfnamefont {A.}~\bibnamefont {Petiteau}}, \bibinfo {author} {\bibfnamefont {A.}~\bibnamefont {Ricciardone}}, \bibinfo {author} {\bibfnamefont {M.}~\bibnamefont {Sakellariadou}}, \bibinfo {author} {\bibfnamefont {L.}~\bibnamefont {Sorbo}},\ and\ \bibinfo {author} {\bibfnamefont {G.}~\bibnamefont {Tasinato}},\ }\bibfield  {title} {\bibinfo {title} {Science with the space-based interferometer
  lisa. iv: probing inflation with gravitational waves},\ }\href {https://doi.org/10.1088/1475-7516/2016/12/026} {\bibfield  {journal} {\bibinfo  {journal} {Journal of Cosmology and Astroparticle Physics}\ }\textbf {\bibinfo {volume} {2016}}\bibinfo  {number} { (12)},\ \bibinfo {pages} {026–026}}\BibitemShut {NoStop}%
\bibitem [{inf(2016)}]{inflation_2016}%
  \BibitemOpen
\bibfield  {number} {  }\href {https://doi.org/10.1393/ncr/i2016-10127-1} {\bibfield  {journal} {\bibinfo  {journal} {La Rivista del Nuovo Cimento}\ }\textbf {\bibinfo {volume} {39}},\ \bibinfo {pages} {399–495} (\bibinfo {year} {2016})}\BibitemShut {NoStop}%
\bibitem [{\citenamefont {Cook}\ and\ \citenamefont {Sorbo}(2012)}]{Cook_2012}%
  \BibitemOpen
  \bibfield  {author} {\bibinfo {author} {\bibfnamefont {J.~L.}\ \bibnamefont {Cook}}\ and\ \bibinfo {author} {\bibfnamefont {L.}~\bibnamefont {Sorbo}},\ }\bibfield  {title} {\bibinfo {title} {Particle production during inflation and gravitational waves detectable by ground-based interferometers},\ }\bibfield  {journal} {\bibinfo  {journal} {Physical Review D}\ }\textbf {\bibinfo {volume} {85}},\ \href {https://doi.org/10.1103/physrevd.85.023534} {10.1103/physrevd.85.023534} (\bibinfo {year} {2012})\BibitemShut {NoStop}%
\bibitem [{\citenamefont {Arvanitaki}\ \emph {et~al.}(2020)\citenamefont {Arvanitaki}, \citenamefont {Dimopoulos}, \citenamefont {Galanis}, \citenamefont {Lehner}, \citenamefont {Thompson},\ and\ \citenamefont {Van~Tilburg}}]{Arvanitaki:2019rax}%
  \BibitemOpen
  \bibfield  {author} {\bibinfo {author} {\bibfnamefont {A.}~\bibnamefont {Arvanitaki}}, \bibinfo {author} {\bibfnamefont {S.}~\bibnamefont {Dimopoulos}}, \bibinfo {author} {\bibfnamefont {M.}~\bibnamefont {Galanis}}, \bibinfo {author} {\bibfnamefont {L.}~\bibnamefont {Lehner}}, \bibinfo {author} {\bibfnamefont {J.~O.}\ \bibnamefont {Thompson}},\ and\ \bibinfo {author} {\bibfnamefont {K.}~\bibnamefont {Van~Tilburg}},\ }\bibfield  {title} {\bibinfo {title} {{Large-misalignment mechanism for the formation of compact axion structures: Signatures from the QCD axion to fuzzy dark matter}},\ }\href {https://doi.org/10.1103/PhysRevD.101.083014} {\bibfield  {journal} {\bibinfo  {journal} {Phys. Rev. D}\ }\textbf {\bibinfo {volume} {101}},\ \bibinfo {pages} {083014} (\bibinfo {year} {2020})},\ \Eprint {https://arxiv.org/abs/1909.11665} {arXiv:1909.11665 [astro-ph.CO]} \BibitemShut {NoStop}%
\bibitem [{\citenamefont {Katz}\ \emph {et~al.}(2021)\citenamefont {Katz} \emph {et~al.}}]{Katz:2021iou}%
  \BibitemOpen
  \bibfield  {author} {\bibinfo {author} {\bibfnamefont {H.}~\bibnamefont {Katz}} \emph {et~al.},\ }\bibfield  {title} {\bibinfo {title} {{Introducing SPHINX-MHD: the impact of primordial magnetic fields on the first galaxies, reionization, and the global 21-cm signal}},\ }\href {https://doi.org/10.1093/mnras/stab2148} {\bibfield  {journal} {\bibinfo  {journal} {Mon. Not. Roy. Astron. Soc.}\ }\textbf {\bibinfo {volume} {507}},\ \bibinfo {pages} {1254} (\bibinfo {year} {2021})},\ \Eprint {https://arxiv.org/abs/2101.11624} {arXiv:2101.11624 [astro-ph.CO]} \BibitemShut {NoStop}%
\bibitem [{\citenamefont {{Ralegankar}}\ \emph {et~al.}(2024)\citenamefont {{Ralegankar}}, \citenamefont {{Pavi{\v{c}}evi{\'c}}},\ and\ \citenamefont {{Viel}}}]{Ralegankar:2024ekl}%
  \BibitemOpen
  \bibfield  {author} {\bibinfo {author} {\bibfnamefont {P.}~\bibnamefont {{Ralegankar}}}, \bibinfo {author} {\bibfnamefont {M.}~\bibnamefont {{Pavi{\v{c}}evi{\'c}}}},\ and\ \bibinfo {author} {\bibfnamefont {M.}~\bibnamefont {{Viel}}},\ }\bibfield  {title} {\bibinfo {title} {{Primordial magnetic fields: consistent initial conditions and impact on high-z structures}},\ }\href {https://doi.org/10.48550/arXiv.2402.14079} {\bibfield  {journal} {\bibinfo  {journal} {arXiv e-prints}\ ,\ \bibinfo {eid} {arXiv:2402.14079}} (\bibinfo {year} {2024})},\ \Eprint {https://arxiv.org/abs/2402.14079} {arXiv:2402.14079 [astro-ph.CO]} \BibitemShut {NoStop}%
\bibitem [{\citenamefont {Lovell}\ \emph {et~al.}(2014)\citenamefont {Lovell}, \citenamefont {Frenk}, \citenamefont {Eke}, \citenamefont {Jenkins}, \citenamefont {Gao},\ and\ \citenamefont {Theuns}}]{Lovell:2013ola}%
  \BibitemOpen
  \bibfield  {author} {\bibinfo {author} {\bibfnamefont {M.~R.}\ \bibnamefont {Lovell}}, \bibinfo {author} {\bibfnamefont {C.~S.}\ \bibnamefont {Frenk}}, \bibinfo {author} {\bibfnamefont {V.~R.}\ \bibnamefont {Eke}}, \bibinfo {author} {\bibfnamefont {A.}~\bibnamefont {Jenkins}}, \bibinfo {author} {\bibfnamefont {L.}~\bibnamefont {Gao}},\ and\ \bibinfo {author} {\bibfnamefont {T.}~\bibnamefont {Theuns}},\ }\bibfield  {title} {\bibinfo {title} {{The properties of warm dark matter haloes}},\ }\href {https://doi.org/10.1093/mnras/stt2431} {\bibfield  {journal} {\bibinfo  {journal} {Mon. Not. Roy. Astron. Soc.}\ }\textbf {\bibinfo {volume} {439}},\ \bibinfo {pages} {300} (\bibinfo {year} {2014})},\ \Eprint {https://arxiv.org/abs/1308.1399} {arXiv:1308.1399 [astro-ph.CO]} \BibitemShut {NoStop}%
\bibitem [{\citenamefont {Boyarsky}\ \emph {et~al.}(2009)\citenamefont {Boyarsky}, \citenamefont {Ruchayskiy},\ and\ \citenamefont {Shaposhnikov}}]{Boyarsky:2009ix}%
  \BibitemOpen
  \bibfield  {author} {\bibinfo {author} {\bibfnamefont {A.}~\bibnamefont {Boyarsky}}, \bibinfo {author} {\bibfnamefont {O.}~\bibnamefont {Ruchayskiy}},\ and\ \bibinfo {author} {\bibfnamefont {M.}~\bibnamefont {Shaposhnikov}},\ }\bibfield  {title} {\bibinfo {title} {{The Role of sterile neutrinos in cosmology and astrophysics}},\ }\href {https://doi.org/10.1146/annurev.nucl.010909.083654} {\bibfield  {journal} {\bibinfo  {journal} {Ann. Rev. Nucl. Part. Sci.}\ }\textbf {\bibinfo {volume} {59}},\ \bibinfo {pages} {191} (\bibinfo {year} {2009})},\ \Eprint {https://arxiv.org/abs/0901.0011} {arXiv:0901.0011 [hep-ph]} \BibitemShut {NoStop}%
\bibitem [{\citenamefont {Egana-Ugrinovic}\ \emph {et~al.}(2021)\citenamefont {Egana-Ugrinovic}, \citenamefont {Essig}, \citenamefont {Gift},\ and\ \citenamefont {LoVerde}}]{Egana-Ugrinovic:2021gnu}%
  \BibitemOpen
  \bibfield  {author} {\bibinfo {author} {\bibfnamefont {D.}~\bibnamefont {Egana-Ugrinovic}}, \bibinfo {author} {\bibfnamefont {R.}~\bibnamefont {Essig}}, \bibinfo {author} {\bibfnamefont {D.}~\bibnamefont {Gift}},\ and\ \bibinfo {author} {\bibfnamefont {M.}~\bibnamefont {LoVerde}},\ }\bibfield  {title} {\bibinfo {title} {{The Cosmological Evolution of Self-interacting Dark Matter}},\ }\href {https://doi.org/10.1088/1475-7516/2021/05/013} {\bibfield  {journal} {\bibinfo  {journal} {JCAP}\ }\textbf {\bibinfo {volume} {05}},\ \bibinfo {pages} {013}},\ \Eprint {https://arxiv.org/abs/2102.06215} {arXiv:2102.06215 [astro-ph.CO]} \BibitemShut {NoStop}%
\bibitem [{\citenamefont {Yi}\ and\ \citenamefont {Fei}(2023)}]{Yi_2023}%
  \BibitemOpen
  \bibfield  {author} {\bibinfo {author} {\bibfnamefont {Z.}~\bibnamefont {Yi}}\ and\ \bibinfo {author} {\bibfnamefont {Q.}~\bibnamefont {Fei}},\ }\bibfield  {title} {\bibinfo {title} {Constraints on primordial curvature spectrum from primordial black holes and scalar-induced gravitational waves},\ }\bibfield  {journal} {\bibinfo  {journal} {The European Physical Journal C}\ }\textbf {\bibinfo {volume} {83}},\ \href {https://doi.org/10.1140/epjc/s10052-023-11233-3} {10.1140/epjc/s10052-023-11233-3} (\bibinfo {year} {2023})\BibitemShut {NoStop}%
\bibitem [{\citenamefont {Braglia}\ \emph {et~al.}(2020)\citenamefont {Braglia}, \citenamefont {Hazra}, \citenamefont {Finelli}, \citenamefont {Smoot}, \citenamefont {Sriramkumar},\ and\ \citenamefont {Starobinsky}}]{Braglia_2020}%
  \BibitemOpen
  \bibfield  {author} {\bibinfo {author} {\bibfnamefont {M.}~\bibnamefont {Braglia}}, \bibinfo {author} {\bibfnamefont {D.~K.}\ \bibnamefont {Hazra}}, \bibinfo {author} {\bibfnamefont {F.}~\bibnamefont {Finelli}}, \bibinfo {author} {\bibfnamefont {G.~F.}\ \bibnamefont {Smoot}}, \bibinfo {author} {\bibfnamefont {L.}~\bibnamefont {Sriramkumar}},\ and\ \bibinfo {author} {\bibfnamefont {A.~A.}\ \bibnamefont {Starobinsky}},\ }\bibfield  {title} {\bibinfo {title} {Generating pbhs and small-scale gws in two-field models of inflation},\ }\href {https://doi.org/10.1088/1475-7516/2020/08/001} {\bibfield  {journal} {\bibinfo  {journal} {Journal of Cosmology and Astroparticle Physics}\ }\textbf {\bibinfo {volume} {2020}}\bibinfo  {number} { (08)},\ \bibinfo {pages} {001–001}}\BibitemShut {NoStop}%
\bibitem [{\citenamefont {Arya}(2020)}]{Arya:2019wck}%
  \BibitemOpen
\bibfield  {number} {  }\bibfield  {author} {\bibinfo {author} {\bibfnamefont {R.}~\bibnamefont {Arya}},\ }\bibfield  {title} {\bibinfo {title} {{Formation of Primordial Black Holes from Warm Inflation}},\ }\href {https://doi.org/10.1088/1475-7516/2020/09/042} {\bibfield  {journal} {\bibinfo  {journal} {JCAP}\ }\textbf {\bibinfo {volume} {09}},\ \bibinfo {pages} {042}},\ \Eprint {https://arxiv.org/abs/1910.05238} {arXiv:1910.05238 [astro-ph.CO]} \BibitemShut {NoStop}%
\bibitem [{\citenamefont {Dom\`enech}(2021)}]{Domenech:2021ztg}%
  \BibitemOpen
  \bibfield  {author} {\bibinfo {author} {\bibfnamefont {G.}~\bibnamefont {Dom\`enech}},\ }\bibfield  {title} {\bibinfo {title} {{Scalar Induced Gravitational Waves Review}},\ }\href {https://doi.org/10.3390/universe7110398} {\bibfield  {journal} {\bibinfo  {journal} {Universe}\ }\textbf {\bibinfo {volume} {7}},\ \bibinfo {pages} {398} (\bibinfo {year} {2021})},\ \Eprint {https://arxiv.org/abs/2109.01398} {arXiv:2109.01398 [gr-qc]} \BibitemShut {NoStop}%
\bibitem [{\citenamefont {Alabidi}\ \emph {et~al.}(2012)\citenamefont {Alabidi}, \citenamefont {Kohri}, \citenamefont {Sasaki},\ and\ \citenamefont {Sendouda}}]{Alabidi:2012ex}%
  \BibitemOpen
  \bibfield  {author} {\bibinfo {author} {\bibfnamefont {L.}~\bibnamefont {Alabidi}}, \bibinfo {author} {\bibfnamefont {K.}~\bibnamefont {Kohri}}, \bibinfo {author} {\bibfnamefont {M.}~\bibnamefont {Sasaki}},\ and\ \bibinfo {author} {\bibfnamefont {Y.}~\bibnamefont {Sendouda}},\ }\bibfield  {title} {\bibinfo {title} {{Observable Spectra of Induced Gravitational Waves from Inflation}},\ }\href {https://doi.org/10.1088/1475-7516/2012/09/017} {\bibfield  {journal} {\bibinfo  {journal} {JCAP}\ }\textbf {\bibinfo {volume} {09}},\ \bibinfo {pages} {017}},\ \Eprint {https://arxiv.org/abs/1203.4663} {arXiv:1203.4663 [astro-ph.CO]} \BibitemShut {NoStop}%
\bibitem [{\citenamefont {Agazie}\ \emph {et~al.}(2023)\citenamefont {Agazie} \emph {et~al.}}]{NANOGrav:2023gor}%
  \BibitemOpen
  \bibfield  {author} {\bibinfo {author} {\bibfnamefont {G.}~\bibnamefont {Agazie}} \emph {et~al.} (\bibinfo {collaboration} {NANOGrav}),\ }\bibfield  {title} {\bibinfo {title} {{The NANOGrav 15 yr Data Set: Evidence for a Gravitational-wave Background}},\ }\href {https://doi.org/10.3847/2041-8213/acdac6} {\bibfield  {journal} {\bibinfo  {journal} {Astrophys. J. Lett.}\ }\textbf {\bibinfo {volume} {951}},\ \bibinfo {pages} {L8} (\bibinfo {year} {2023})},\ \Eprint {https://arxiv.org/abs/2306.16213} {arXiv:2306.16213 [astro-ph.HE]} \BibitemShut {NoStop}%
\bibitem [{\citenamefont {Antoniadis}\ \emph {et~al.}(2023)\citenamefont {Antoniadis} \emph {et~al.}}]{EPTA:2023fyk}%
  \BibitemOpen
  \bibfield  {author} {\bibinfo {author} {\bibfnamefont {J.}~\bibnamefont {Antoniadis}} \emph {et~al.} (\bibinfo {collaboration} {EPTA, InPTA:}),\ }\bibfield  {title} {\bibinfo {title} {{The second data release from the European Pulsar Timing Array - III. Search for gravitational wave signals}},\ }\href {https://doi.org/10.1051/0004-6361/202346844} {\bibfield  {journal} {\bibinfo  {journal} {Astron. Astrophys.}\ }\textbf {\bibinfo {volume} {678}},\ \bibinfo {pages} {A50} (\bibinfo {year} {2023})},\ \Eprint {https://arxiv.org/abs/2306.16214} {arXiv:2306.16214 [astro-ph.HE]} \BibitemShut {NoStop}%
\bibitem [{\citenamefont {Xu}\ \emph {et~al.}(2023)\citenamefont {Xu} \emph {et~al.}}]{Xu:2023wog}%
  \BibitemOpen
  \bibfield  {author} {\bibinfo {author} {\bibfnamefont {H.}~\bibnamefont {Xu}} \emph {et~al.},\ }\bibfield  {title} {\bibinfo {title} {{Searching for the Nano-Hertz Stochastic Gravitational Wave Background with the Chinese Pulsar Timing Array Data Release I}},\ }\href {https://doi.org/10.1088/1674-4527/acdfa5} {\bibfield  {journal} {\bibinfo  {journal} {Res. Astron. Astrophys.}\ }\textbf {\bibinfo {volume} {23}},\ \bibinfo {pages} {075024} (\bibinfo {year} {2023})},\ \Eprint {https://arxiv.org/abs/2306.16216} {arXiv:2306.16216 [astro-ph.HE]} \BibitemShut {NoStop}%
\bibitem [{\citenamefont {Reardon}\ \emph {et~al.}(2023)\citenamefont {Reardon} \emph {et~al.}}]{Reardon:2023gzh}%
  \BibitemOpen
  \bibfield  {author} {\bibinfo {author} {\bibfnamefont {D.~J.}\ \bibnamefont {Reardon}} \emph {et~al.},\ }\bibfield  {title} {\bibinfo {title} {{Search for an Isotropic Gravitational-wave Background with the Parkes Pulsar Timing Array}},\ }\href {https://doi.org/10.3847/2041-8213/acdd02} {\bibfield  {journal} {\bibinfo  {journal} {Astrophys. J. Lett.}\ }\textbf {\bibinfo {volume} {951}},\ \bibinfo {pages} {L6} (\bibinfo {year} {2023})},\ \Eprint {https://arxiv.org/abs/2306.16215} {arXiv:2306.16215 [astro-ph.HE]} \BibitemShut {NoStop}%
\bibitem [{\citenamefont {Afzal}\ \emph {et~al.}(2023)\citenamefont {Afzal} \emph {et~al.}}]{Afzal_2023}%
  \BibitemOpen
  \bibfield  {author} {\bibinfo {author} {\bibfnamefont {A.}~\bibnamefont {Afzal}} \emph {et~al.} (\bibinfo {collaboration} {The NANOGrav Collaboration}),\ }\bibfield  {title} {\bibinfo {title} {The nanograv 15 yr data set: Search for signals from new physics},\ }\href {https://doi.org/10.3847/2041-8213/acdc91} {\bibfield  {journal} {\bibinfo  {journal} {The Astrophysical Journal Letters}\ }\textbf {\bibinfo {volume} {951}},\ \bibinfo {pages} {L11} (\bibinfo {year} {2023})}\BibitemShut {NoStop}%
\bibitem [{\citenamefont {Nakama}\ \emph {et~al.}(2017)\citenamefont {Nakama}, \citenamefont {Chluba},\ and\ \citenamefont {Kamionkowski}}]{Nakama:2017ohe}%
  \BibitemOpen
  \bibfield  {author} {\bibinfo {author} {\bibfnamefont {T.}~\bibnamefont {Nakama}}, \bibinfo {author} {\bibfnamefont {J.}~\bibnamefont {Chluba}},\ and\ \bibinfo {author} {\bibfnamefont {M.}~\bibnamefont {Kamionkowski}},\ }\bibfield  {title} {\bibinfo {title} {{Shedding light on the small-scale crisis with CMB spectral distortions}},\ }\href {https://doi.org/10.1103/PhysRevD.95.121302} {\bibfield  {journal} {\bibinfo  {journal} {Phys. Rev. D}\ }\textbf {\bibinfo {volume} {95}},\ \bibinfo {pages} {121302} (\bibinfo {year} {2017})},\ \Eprint {https://arxiv.org/abs/1703.10559} {arXiv:1703.10559 [astro-ph.CO]} \BibitemShut {NoStop}%
\bibitem [{\citenamefont {Kogut}\ \emph {et~al.}(2019)\citenamefont {Kogut}, \citenamefont {Abitbol}, \citenamefont {Chluba}, \citenamefont {Delabrouille}, \citenamefont {Fixsen}, \citenamefont {Hill}, \citenamefont {Patil},\ and\ \citenamefont {Rotti}}]{Kogut:2019vqh}%
  \BibitemOpen
  \bibfield  {author} {\bibinfo {author} {\bibfnamefont {A.}~\bibnamefont {Kogut}}, \bibinfo {author} {\bibfnamefont {M.~H.}\ \bibnamefont {Abitbol}}, \bibinfo {author} {\bibfnamefont {J.}~\bibnamefont {Chluba}}, \bibinfo {author} {\bibfnamefont {J.}~\bibnamefont {Delabrouille}}, \bibinfo {author} {\bibfnamefont {D.}~\bibnamefont {Fixsen}}, \bibinfo {author} {\bibfnamefont {J.~C.}\ \bibnamefont {Hill}}, \bibinfo {author} {\bibfnamefont {S.~P.}\ \bibnamefont {Patil}},\ and\ \bibinfo {author} {\bibfnamefont {A.}~\bibnamefont {Rotti}},\ }\bibfield  {title} {\bibinfo {title} {{CMB Spectral Distortions: Status and Prospects}},\ }\href@noop {} {\bibfield  {journal} {\bibinfo  {journal} {Bull. Am. Astron. Soc.}\ }\textbf {\bibinfo {volume} {51}},\ \bibinfo {pages} {113} (\bibinfo {year} {2019})},\ \Eprint {https://arxiv.org/abs/1907.13195} {arXiv:1907.13195 [astro-ph.CO]} \BibitemShut {NoStop}%
\bibitem [{\citenamefont {Kennedy}\ \emph {et~al.}(2014)\citenamefont {Kennedy}, \citenamefont {Frenk}, \citenamefont {Cole},\ and\ \citenamefont {Benson}}]{Kennedy_2014}%
  \BibitemOpen
  \bibfield  {author} {\bibinfo {author} {\bibfnamefont {R.}~\bibnamefont {Kennedy}}, \bibinfo {author} {\bibfnamefont {C.}~\bibnamefont {Frenk}}, \bibinfo {author} {\bibfnamefont {S.}~\bibnamefont {Cole}},\ and\ \bibinfo {author} {\bibfnamefont {A.}~\bibnamefont {Benson}},\ }\bibfield  {title} {\bibinfo {title} {Constraining the warm dark matter particle mass with milky way satellites},\ }\href {https://doi.org/10.1093/mnras/stu719} {\bibfield  {journal} {\bibinfo  {journal} {Monthly Notices of the Royal Astronomical Society}\ }\textbf {\bibinfo {volume} {442}},\ \bibinfo {pages} {2487–2495} (\bibinfo {year} {2014})}\BibitemShut {NoStop}%
\bibitem [{\citenamefont {Nadler}\ \emph {et~al.}(2021)\citenamefont {Nadler} \emph {et~al.}}]{DES:2020fxi}%
  \BibitemOpen
  \bibfield  {author} {\bibinfo {author} {\bibfnamefont {E.~O.}\ \bibnamefont {Nadler}} \emph {et~al.} (\bibinfo {collaboration} {DES}),\ }\bibfield  {title} {\bibinfo {title} {{Milky Way Satellite Census. III. Constraints on Dark Matter Properties from Observations of Milky Way Satellite Galaxies}},\ }\href {https://doi.org/10.1103/PhysRevLett.126.091101} {\bibfield  {journal} {\bibinfo  {journal} {Phys. Rev. Lett.}\ }\textbf {\bibinfo {volume} {126}},\ \bibinfo {pages} {091101} (\bibinfo {year} {2021})},\ \Eprint {https://arxiv.org/abs/2008.00022} {arXiv:2008.00022 [astro-ph.CO]} \BibitemShut {NoStop}%
\bibitem [{\citenamefont {{Nadler}}\ \emph {et~al.}(2021)\citenamefont {{Nadler}}, \citenamefont {{Drlica-Wagner}}, \citenamefont {{Bechtol}}, \citenamefont {{Mau}}, \citenamefont {{Wechsler}}, \citenamefont {{Gluscevic}}, \citenamefont {{Boddy}}, \citenamefont {{Pace}}, \citenamefont {{Li}}, \citenamefont {{McNanna}}, \citenamefont {{Riley}}, \citenamefont {{Garc{\'\i}a-Bellido}}, \citenamefont {{Mao}}, \citenamefont {{Green}}, \citenamefont {{Burke}}, \citenamefont {{Peter}}, \citenamefont {{Jain}}, \citenamefont {{Abbott}}, \citenamefont {{Aguena}}, \citenamefont {{Allam}}, \citenamefont {{Annis}}, \citenamefont {{Avila}}, \citenamefont {{Brooks}}, \citenamefont {{Carrasco Kind}}, \citenamefont {{Carretero}}, \citenamefont {{Costanzi}}, \citenamefont {{da Costa}}, \citenamefont {{De Vicente}}, \citenamefont {{Desai}}, \citenamefont {{Diehl}}, \citenamefont {{Doel}}, \citenamefont {{Everett}}, \citenamefont {{Evrard}}, \citenamefont {{Flaugher}}, \citenamefont {{Frieman}}, \citenamefont {{Gerdes}},
  \citenamefont {{Gruen}}, \citenamefont {{Gruendl}}, \citenamefont {{Gschwend}}, \citenamefont {{Gutierrez}}, \citenamefont {{Hinton}}, \citenamefont {{Honscheid}}, \citenamefont {{Huterer}}, \citenamefont {{James}}, \citenamefont {{Krause}}, \citenamefont {{Kuehn}}, \citenamefont {{Kuropatkin}}, \citenamefont {{Lahav}}, \citenamefont {{Maia}}, \citenamefont {{Marshall}}, \citenamefont {{Menanteau}}, \citenamefont {{Miquel}}, \citenamefont {{Palmese}}, \citenamefont {{Paz-Chinch{\'o}n}}, \citenamefont {{Plazas}}, \citenamefont {{Romer}}, \citenamefont {{Sanchez}}, \citenamefont {{Scarpine}}, \citenamefont {{Serrano}}, \citenamefont {{Sevilla-Noarbe}}, \citenamefont {{Smith}}, \citenamefont {{Soares-Santos}}, \citenamefont {{Suchyta}}, \citenamefont {{Swanson}}, \citenamefont {{Tarle}}, \citenamefont {{Tucker}}, \citenamefont {{Walker}}, \citenamefont {{Wester}},\ and\ \citenamefont {{DES Collaboration}}}]{Nadler.etal.2021}%
  \BibitemOpen
  \bibfield  {author} {\bibinfo {author} {\bibfnamefont {E.~O.}\ \bibnamefont {{Nadler}}}, \bibinfo {author} {\bibfnamefont {A.}~\bibnamefont {{Drlica-Wagner}}}, \bibinfo {author} {\bibfnamefont {K.}~\bibnamefont {{Bechtol}}}, \bibinfo {author} {\bibfnamefont {S.}~\bibnamefont {{Mau}}}, \bibinfo {author} {\bibfnamefont {R.~H.}\ \bibnamefont {{Wechsler}}}, \bibinfo {author} {\bibfnamefont {V.}~\bibnamefont {{Gluscevic}}}, \bibinfo {author} {\bibfnamefont {K.}~\bibnamefont {{Boddy}}}, \bibinfo {author} {\bibfnamefont {A.~B.}\ \bibnamefont {{Pace}}}, \bibinfo {author} {\bibfnamefont {T.~S.}\ \bibnamefont {{Li}}}, \bibinfo {author} {\bibfnamefont {M.}~\bibnamefont {{McNanna}}}, \bibinfo {author} {\bibfnamefont {A.~H.}\ \bibnamefont {{Riley}}}, \bibinfo {author} {\bibfnamefont {J.}~\bibnamefont {{Garc{\'\i}a-Bellido}}}, \bibinfo {author} {\bibfnamefont {Y.~Y.}\ \bibnamefont {{Mao}}}, \bibinfo {author} {\bibfnamefont {G.}~\bibnamefont {{Green}}}, \bibinfo {author} {\bibfnamefont {D.~L.}\ \bibnamefont {{Burke}}},
  \bibinfo {author} {\bibfnamefont {A.}~\bibnamefont {{Peter}}}, \bibinfo {author} {\bibfnamefont {B.}~\bibnamefont {{Jain}}}, \bibinfo {author} {\bibfnamefont {T.~M.~C.}\ \bibnamefont {{Abbott}}}, \bibinfo {author} {\bibfnamefont {M.}~\bibnamefont {{Aguena}}}, \bibinfo {author} {\bibfnamefont {S.}~\bibnamefont {{Allam}}}, \bibinfo {author} {\bibfnamefont {J.}~\bibnamefont {{Annis}}}, \bibinfo {author} {\bibfnamefont {S.}~\bibnamefont {{Avila}}}, \bibinfo {author} {\bibfnamefont {D.}~\bibnamefont {{Brooks}}}, \bibinfo {author} {\bibfnamefont {M.}~\bibnamefont {{Carrasco Kind}}}, \bibinfo {author} {\bibfnamefont {J.}~\bibnamefont {{Carretero}}}, \bibinfo {author} {\bibfnamefont {M.}~\bibnamefont {{Costanzi}}}, \bibinfo {author} {\bibfnamefont {L.~N.}\ \bibnamefont {{da Costa}}}, \bibinfo {author} {\bibfnamefont {J.}~\bibnamefont {{De Vicente}}}, \bibinfo {author} {\bibfnamefont {S.}~\bibnamefont {{Desai}}}, \bibinfo {author} {\bibfnamefont {H.~T.}\ \bibnamefont {{Diehl}}}, \bibinfo {author} {\bibfnamefont
  {P.}~\bibnamefont {{Doel}}}, \bibinfo {author} {\bibfnamefont {S.}~\bibnamefont {{Everett}}}, \bibinfo {author} {\bibfnamefont {A.~E.}\ \bibnamefont {{Evrard}}}, \bibinfo {author} {\bibfnamefont {B.}~\bibnamefont {{Flaugher}}}, \bibinfo {author} {\bibfnamefont {J.}~\bibnamefont {{Frieman}}}, \bibinfo {author} {\bibfnamefont {D.~W.}\ \bibnamefont {{Gerdes}}}, \bibinfo {author} {\bibfnamefont {D.}~\bibnamefont {{Gruen}}}, \bibinfo {author} {\bibfnamefont {R.~A.}\ \bibnamefont {{Gruendl}}}, \bibinfo {author} {\bibfnamefont {J.}~\bibnamefont {{Gschwend}}}, \bibinfo {author} {\bibfnamefont {G.}~\bibnamefont {{Gutierrez}}}, \bibinfo {author} {\bibfnamefont {S.~R.}\ \bibnamefont {{Hinton}}}, \bibinfo {author} {\bibfnamefont {K.}~\bibnamefont {{Honscheid}}}, \bibinfo {author} {\bibfnamefont {D.}~\bibnamefont {{Huterer}}}, \bibinfo {author} {\bibfnamefont {D.~J.}\ \bibnamefont {{James}}}, \bibinfo {author} {\bibfnamefont {E.}~\bibnamefont {{Krause}}}, \bibinfo {author} {\bibfnamefont {K.}~\bibnamefont {{Kuehn}}},
  \bibinfo {author} {\bibfnamefont {N.}~\bibnamefont {{Kuropatkin}}}, \bibinfo {author} {\bibfnamefont {O.}~\bibnamefont {{Lahav}}}, \bibinfo {author} {\bibfnamefont {M.~A.~G.}\ \bibnamefont {{Maia}}}, \bibinfo {author} {\bibfnamefont {J.~L.}\ \bibnamefont {{Marshall}}}, \bibinfo {author} {\bibfnamefont {F.}~\bibnamefont {{Menanteau}}}, \bibinfo {author} {\bibfnamefont {R.}~\bibnamefont {{Miquel}}}, \bibinfo {author} {\bibfnamefont {A.}~\bibnamefont {{Palmese}}}, \bibinfo {author} {\bibfnamefont {F.}~\bibnamefont {{Paz-Chinch{\'o}n}}}, \bibinfo {author} {\bibfnamefont {A.~A.}\ \bibnamefont {{Plazas}}}, \bibinfo {author} {\bibfnamefont {A.~K.}\ \bibnamefont {{Romer}}}, \bibinfo {author} {\bibfnamefont {E.}~\bibnamefont {{Sanchez}}}, \bibinfo {author} {\bibfnamefont {V.}~\bibnamefont {{Scarpine}}}, \bibinfo {author} {\bibfnamefont {S.}~\bibnamefont {{Serrano}}}, \bibinfo {author} {\bibfnamefont {I.}~\bibnamefont {{Sevilla-Noarbe}}}, \bibinfo {author} {\bibfnamefont {M.}~\bibnamefont {{Smith}}}, \bibinfo
  {author} {\bibfnamefont {M.}~\bibnamefont {{Soares-Santos}}}, \bibinfo {author} {\bibfnamefont {E.}~\bibnamefont {{Suchyta}}}, \bibinfo {author} {\bibfnamefont {M.~E.~C.}\ \bibnamefont {{Swanson}}}, \bibinfo {author} {\bibfnamefont {G.}~\bibnamefont {{Tarle}}}, \bibinfo {author} {\bibfnamefont {D.~L.}\ \bibnamefont {{Tucker}}}, \bibinfo {author} {\bibfnamefont {A.~R.}\ \bibnamefont {{Walker}}}, \bibinfo {author} {\bibfnamefont {W.}~\bibnamefont {{Wester}}},\ and\ \bibinfo {author} {\bibnamefont {{DES Collaboration}}},\ }\bibfield  {title} {\bibinfo {title} {{Constraints on Dark Matter Properties from Observations of Milky Way Satellite Galaxies}},\ }\href {https://doi.org/10.1103/PhysRevLett.126.091101} {\bibfield  {journal} {\bibinfo  {journal} {\prl}\ }\textbf {\bibinfo {volume} {126}},\ \bibinfo {eid} {091101} (\bibinfo {year} {2021})},\ \Eprint {https://arxiv.org/abs/2008.00022} {arXiv:2008.00022 [astro-ph.CO]} \BibitemShut {NoStop}%
\bibitem [{\citenamefont {Kim}\ and\ \citenamefont {Peter}(2022)}]{kim2022milky}%
  \BibitemOpen
  \bibfield  {author} {\bibinfo {author} {\bibfnamefont {S.~Y.}\ \bibnamefont {Kim}}\ and\ \bibinfo {author} {\bibfnamefont {A.~H.~G.}\ \bibnamefont {Peter}},\ }\href@noop {} {\bibinfo {title} {The milky way satellite velocity function is a sharp probe of small-scale structure problems}} (\bibinfo {year} {2022}),\ \Eprint {https://arxiv.org/abs/2106.09050} {arXiv:2106.09050 [astro-ph.GA]} \BibitemShut {NoStop}%
\bibitem [{\citenamefont {Cherry}\ and\ \citenamefont {Horiuchi}(2017)}]{PhysRevD.95.083015}%
  \BibitemOpen
  \bibfield  {author} {\bibinfo {author} {\bibfnamefont {J.~F.}\ \bibnamefont {Cherry}}\ and\ \bibinfo {author} {\bibfnamefont {S.}~\bibnamefont {Horiuchi}},\ }\bibfield  {title} {\bibinfo {title} {Closing in on resonantly produced sterile neutrino dark matter},\ }\href {https://doi.org/10.1103/PhysRevD.95.083015} {\bibfield  {journal} {\bibinfo  {journal} {Phys. Rev. D}\ }\textbf {\bibinfo {volume} {95}},\ \bibinfo {pages} {083015} (\bibinfo {year} {2017})}\BibitemShut {NoStop}%
\bibitem [{\citenamefont {Newton}\ \emph {et~al.}(2021)\citenamefont {Newton}, \citenamefont {Leo}, \citenamefont {Cautun}, \citenamefont {Jenkins}, \citenamefont {Frenk}, \citenamefont {Lovell}, \citenamefont {Helly}, \citenamefont {Benson},\ and\ \citenamefont {Cole}}]{Newton:2020cog}%
  \BibitemOpen
  \bibfield  {author} {\bibinfo {author} {\bibfnamefont {O.}~\bibnamefont {Newton}}, \bibinfo {author} {\bibfnamefont {M.}~\bibnamefont {Leo}}, \bibinfo {author} {\bibfnamefont {M.}~\bibnamefont {Cautun}}, \bibinfo {author} {\bibfnamefont {A.}~\bibnamefont {Jenkins}}, \bibinfo {author} {\bibfnamefont {C.~S.}\ \bibnamefont {Frenk}}, \bibinfo {author} {\bibfnamefont {M.~R.}\ \bibnamefont {Lovell}}, \bibinfo {author} {\bibfnamefont {J.~C.}\ \bibnamefont {Helly}}, \bibinfo {author} {\bibfnamefont {A.~J.}\ \bibnamefont {Benson}},\ and\ \bibinfo {author} {\bibfnamefont {S.}~\bibnamefont {Cole}},\ }\bibfield  {title} {\bibinfo {title} {{Constraints on the properties of warm dark matter using the satellite galaxies of the Milky Way}},\ }\href {https://doi.org/10.1088/1475-7516/2021/08/062} {\bibfield  {journal} {\bibinfo  {journal} {JCAP}\ }\textbf {\bibinfo {volume} {08}},\ \bibinfo {pages} {062}},\ \Eprint {https://arxiv.org/abs/2011.08865} {arXiv:2011.08865 [astro-ph.CO]} \BibitemShut {NoStop}%
\bibitem [{\citenamefont {Dekker}\ \emph {et~al.}(2022)\citenamefont {Dekker}, \citenamefont {Ando}, \citenamefont {Correa},\ and\ \citenamefont {Ng}}]{Dekker_2022}%
  \BibitemOpen
  \bibfield  {author} {\bibinfo {author} {\bibfnamefont {A.}~\bibnamefont {Dekker}}, \bibinfo {author} {\bibfnamefont {S.}~\bibnamefont {Ando}}, \bibinfo {author} {\bibfnamefont {C.~A.}\ \bibnamefont {Correa}},\ and\ \bibinfo {author} {\bibfnamefont {K.~C.}\ \bibnamefont {Ng}},\ }\bibfield  {title} {\bibinfo {title} {Warm dark matter constraints using milky-way satellite observations and subhalo evolution modeling},\ }\bibfield  {journal} {\bibinfo  {journal} {Physical Review D}\ }\textbf {\bibinfo {volume} {106}},\ \href {https://doi.org/10.1103/physrevd.106.123026} {10.1103/physrevd.106.123026} (\bibinfo {year} {2022})\BibitemShut {NoStop}%
\bibitem [{\citenamefont {{Dalal}}\ and\ \citenamefont {{Kravtsov}}(2022)}]{Dalal.Kravtsov.2022}%
  \BibitemOpen
  \bibfield  {author} {\bibinfo {author} {\bibfnamefont {N.}~\bibnamefont {{Dalal}}}\ and\ \bibinfo {author} {\bibfnamefont {A.}~\bibnamefont {{Kravtsov}}},\ }\bibfield  {title} {\bibinfo {title} {{Excluding fuzzy dark matter with sizes and stellar kinematics of ultrafaint dwarf galaxies}},\ }\href {https://doi.org/10.1103/PhysRevD.106.063517} {\bibfield  {journal} {\bibinfo  {journal} {\prd}\ }\textbf {\bibinfo {volume} {106}},\ \bibinfo {eid} {063517} (\bibinfo {year} {2022})}\BibitemShut {NoStop}%
\bibitem [{\citenamefont {Gilman}\ \emph {et~al.}(2022)\citenamefont {Gilman}, \citenamefont {Benson}, \citenamefont {Bovy}, \citenamefont {Birrer}, \citenamefont {Treu},\ and\ \citenamefont {Nierenberg}}]{Gilman_2022}%
  \BibitemOpen
  \bibfield  {author} {\bibinfo {author} {\bibfnamefont {D.}~\bibnamefont {Gilman}}, \bibinfo {author} {\bibfnamefont {A.}~\bibnamefont {Benson}}, \bibinfo {author} {\bibfnamefont {J.}~\bibnamefont {Bovy}}, \bibinfo {author} {\bibfnamefont {S.}~\bibnamefont {Birrer}}, \bibinfo {author} {\bibfnamefont {T.}~\bibnamefont {Treu}},\ and\ \bibinfo {author} {\bibfnamefont {A.}~\bibnamefont {Nierenberg}},\ }\bibfield  {title} {\bibinfo {title} {The primordial matter power spectrum on sub-galactic scales},\ }\href {https://doi.org/10.1093/mnras/stac670} {\bibfield  {journal} {\bibinfo  {journal} {Monthly Notices of the Royal Astronomical Society}\ }\textbf {\bibinfo {volume} {512}},\ \bibinfo {pages} {3163–3188} (\bibinfo {year} {2022})}\BibitemShut {NoStop}%
\bibitem [{\citenamefont {Esteban}\ \emph {et~al.}(2023)\citenamefont {Esteban}, \citenamefont {Peter},\ and\ \citenamefont {Kim}}]{esteban2023milky}%
  \BibitemOpen
  \bibfield  {author} {\bibinfo {author} {\bibfnamefont {I.}~\bibnamefont {Esteban}}, \bibinfo {author} {\bibfnamefont {A.~H.~G.}\ \bibnamefont {Peter}},\ and\ \bibinfo {author} {\bibfnamefont {S.~Y.}\ \bibnamefont {Kim}},\ }\href@noop {} {\bibinfo {title} {Milky way satellite velocities reveal the dark matter power spectrum at small scales}} (\bibinfo {year} {2023}),\ \Eprint {https://arxiv.org/abs/2306.04674} {arXiv:2306.04674 [astro-ph.CO]} \BibitemShut {NoStop}%
\bibitem [{\citenamefont {Kravtsov}\ and\ \citenamefont {Wu}(2023)}]{Kravtsov_2023}%
  \BibitemOpen
  \bibfield  {author} {\bibinfo {author} {\bibfnamefont {A.}~\bibnamefont {Kravtsov}}\ and\ \bibinfo {author} {\bibfnamefont {Z.}~\bibnamefont {Wu}},\ }\bibfield  {title} {\bibinfo {title} {Densities and mass assembly histories of the milky way satellites are not a challenge to $\lambda$cdm},\ }\href {https://doi.org/10.1093/mnras/stad2219} {\bibfield  {journal} {\bibinfo  {journal} {Monthly Notices of the Royal Astronomical Society}\ }\textbf {\bibinfo {volume} {525}},\ \bibinfo {pages} {325–334} (\bibinfo {year} {2023})}\BibitemShut {NoStop}%
\bibitem [{\citenamefont {{Kravtsov}}\ and\ \citenamefont {{Manwadkar}}(2022)}]{2022MNRAS.514.2667K}%
  \BibitemOpen
  \bibfield  {author} {\bibinfo {author} {\bibfnamefont {A.}~\bibnamefont {{Kravtsov}}}\ and\ \bibinfo {author} {\bibfnamefont {V.}~\bibnamefont {{Manwadkar}}},\ }\bibfield  {title} {\bibinfo {title} {{GRUMPY: a simple framework for realistic forward modelling of dwarf galaxies}},\ }\href {https://doi.org/10.1093/mnras/stac1439} {\bibfield  {journal} {\bibinfo  {journal} {\mnras}\ }\textbf {\bibinfo {volume} {514}},\ \bibinfo {pages} {2667} (\bibinfo {year} {2022})},\ \Eprint {https://arxiv.org/abs/2106.09724} {arXiv:2106.09724 [astro-ph.GA]} \BibitemShut {NoStop}%
\bibitem [{\citenamefont {Simon}(2019)}]{Simon:2019nxf}%
  \BibitemOpen
  \bibfield  {author} {\bibinfo {author} {\bibfnamefont {J.~D.}\ \bibnamefont {Simon}},\ }\bibfield  {title} {\bibinfo {title} {{The Faintest Dwarf Galaxies}},\ }\href {https://doi.org/10.1146/annurev-astro-091918-104453} {\bibfield  {journal} {\bibinfo  {journal} {Ann. Rev. Astron. Astrophys.}\ }\textbf {\bibinfo {volume} {57}},\ \bibinfo {pages} {375} (\bibinfo {year} {2019})},\ \Eprint {https://arxiv.org/abs/1901.05465} {arXiv:1901.05465 [astro-ph.GA]} \BibitemShut {NoStop}%
\bibitem [{\citenamefont {{Kravtsov}}(2024)}]{Kravtsov:2024esa}%
  \BibitemOpen
  \bibfield  {author} {\bibinfo {author} {\bibfnamefont {A.}~\bibnamefont {{Kravtsov}}},\ }\bibfield  {title} {\bibinfo {title} {{On the dark matter content of ultra-diffuse galaxies}},\ }\href {https://doi.org/10.48550/arXiv.2406.13732} {\bibfield  {journal} {\bibinfo  {journal} {arXiv e-prints}\ ,\ \bibinfo {eid} {arXiv:2406.13732}} (\bibinfo {year} {2024})},\ \Eprint {https://arxiv.org/abs/2406.13732} {arXiv:2406.13732 [astro-ph.GA]} \BibitemShut {NoStop}%
\bibitem [{\citenamefont {Wolf}\ \emph {et~al.}(2010)\citenamefont {Wolf}, \citenamefont {Martinez}, \citenamefont {Bullock}, \citenamefont {Kaplinghat}, \citenamefont {Geha}, \citenamefont {Muñoz}, \citenamefont {Simon},\ and\ \citenamefont {Avedo}}]{Wolf_2010}%
  \BibitemOpen
  \bibfield  {author} {\bibinfo {author} {\bibfnamefont {J.}~\bibnamefont {Wolf}}, \bibinfo {author} {\bibfnamefont {G.~D.}\ \bibnamefont {Martinez}}, \bibinfo {author} {\bibfnamefont {J.~S.}\ \bibnamefont {Bullock}}, \bibinfo {author} {\bibfnamefont {M.}~\bibnamefont {Kaplinghat}}, \bibinfo {author} {\bibfnamefont {M.}~\bibnamefont {Geha}}, \bibinfo {author} {\bibfnamefont {R.~R.}\ \bibnamefont {Muñoz}}, \bibinfo {author} {\bibfnamefont {J.~D.}\ \bibnamefont {Simon}},\ and\ \bibinfo {author} {\bibfnamefont {F.~F.}\ \bibnamefont {Avedo}},\ }\bibfield  {title} {\bibinfo {title} {Accurate masses for dispersion-supported galaxies: Accurate masses for spheroidal galaxies},\ }\href {https://doi.org/10.1111/j.1365-2966.2010.16753.x} {\bibfield  {journal} {\bibinfo  {journal} {Monthly Notices of the Royal Astronomical Society}\ ,\ \bibinfo {pages} {no}} (\bibinfo {year} {2010})}\BibitemShut {NoStop}%
\bibitem [{\citenamefont {Chan}\ \emph {et~al.}(2017)\citenamefont {Chan}, \citenamefont {Sheth},\ and\ \citenamefont {Scoccimarro}}]{Chan:2015zjt}%
  \BibitemOpen
  \bibfield  {author} {\bibinfo {author} {\bibfnamefont {K.~C.}\ \bibnamefont {Chan}}, \bibinfo {author} {\bibfnamefont {R.~K.}\ \bibnamefont {Sheth}},\ and\ \bibinfo {author} {\bibfnamefont {R.}~\bibnamefont {Scoccimarro}},\ }\bibfield  {title} {\bibinfo {title} {{Effective Window Function for Lagrangian Halos}},\ }\href {https://doi.org/10.1103/PhysRevD.96.103543} {\bibfield  {journal} {\bibinfo  {journal} {Phys. Rev. D}\ }\textbf {\bibinfo {volume} {96}},\ \bibinfo {pages} {103543} (\bibinfo {year} {2017})},\ \Eprint {https://arxiv.org/abs/1511.01909} {arXiv:1511.01909 [astro-ph.CO]} \BibitemShut {NoStop}%
\bibitem [{\citenamefont {{Diemer}}\ and\ \citenamefont {{Joyce}}(2019)}]{Diemer.Joyce.2019}%
  \BibitemOpen
  \bibfield  {author} {\bibinfo {author} {\bibfnamefont {B.}~\bibnamefont {{Diemer}}}\ and\ \bibinfo {author} {\bibfnamefont {M.}~\bibnamefont {{Joyce}}},\ }\bibfield  {title} {\bibinfo {title} {{An Accurate Physical Model for Halo Concentrations}},\ }\href {https://doi.org/10.3847/1538-4357/aafad6} {\bibfield  {journal} {\bibinfo  {journal} {\apj}\ }\textbf {\bibinfo {volume} {871}},\ \bibinfo {eid} {168} (\bibinfo {year} {2019})},\ \Eprint {https://arxiv.org/abs/1809.07326} {arXiv:1809.07326 [astro-ph.CO]} \BibitemShut {NoStop}%
\bibitem [{\citenamefont {Parashari}\ and\ \citenamefont {Laha}(2023)}]{Parashari_2023}%
  \BibitemOpen
  \bibfield  {author} {\bibinfo {author} {\bibfnamefont {P.}~\bibnamefont {Parashari}}\ and\ \bibinfo {author} {\bibfnamefont {R.}~\bibnamefont {Laha}},\ }\bibfield  {title} {\bibinfo {title} {Primordial power spectrum in light of jwst observations of high redshift galaxies},\ }\href {https://doi.org/10.1093/mnrasl/slad107} {\bibfield  {journal} {\bibinfo  {journal} {Monthly Notices of the Royal Astronomical Society: Letters}\ }\textbf {\bibinfo {volume} {526}},\ \bibinfo {pages} {L63–L69} (\bibinfo {year} {2023})}\BibitemShut {NoStop}%
\bibitem [{\citenamefont {Hirano}\ \emph {et~al.}(2015)\citenamefont {Hirano}, \citenamefont {Zhu}, \citenamefont {Yoshida}, \citenamefont {Spergel},\ and\ \citenamefont {Yorke}}]{Hirano:2015wla}%
  \BibitemOpen
  \bibfield  {author} {\bibinfo {author} {\bibfnamefont {S.}~\bibnamefont {Hirano}}, \bibinfo {author} {\bibfnamefont {N.}~\bibnamefont {Zhu}}, \bibinfo {author} {\bibfnamefont {N.}~\bibnamefont {Yoshida}}, \bibinfo {author} {\bibfnamefont {D.}~\bibnamefont {Spergel}},\ and\ \bibinfo {author} {\bibfnamefont {H.~W.}\ \bibnamefont {Yorke}},\ }\bibfield  {title} {\bibinfo {title} {{Early structure formation from primordial density fluctuations with a blue-tilted power spectrum}},\ }\href {https://doi.org/10.1088/0004-637X/814/1/18} {\bibfield  {journal} {\bibinfo  {journal} {Astrophys. J.}\ }\textbf {\bibinfo {volume} {814}},\ \bibinfo {pages} {18} (\bibinfo {year} {2015})},\ \Eprint {https://arxiv.org/abs/1504.05186} {arXiv:1504.05186 [astro-ph.CO]} \BibitemShut {NoStop}%
\bibitem [{\citenamefont {{Eisenstein}}\ and\ \citenamefont {{Hu}}(1998)}]{1998ApJ...496..605E}%
  \BibitemOpen
  \bibfield  {author} {\bibinfo {author} {\bibfnamefont {D.~J.}\ \bibnamefont {{Eisenstein}}}\ and\ \bibinfo {author} {\bibfnamefont {W.}~\bibnamefont {{Hu}}},\ }\bibfield  {title} {\bibinfo {title} {{Baryonic Features in the Matter Transfer Function}},\ }\href {https://doi.org/10.1086/305424} {\bibfield  {journal} {\bibinfo  {journal} {\apj}\ }\textbf {\bibinfo {volume} {496}},\ \bibinfo {pages} {605} (\bibinfo {year} {1998})},\ \Eprint {https://arxiv.org/abs/astro-ph/9709112} {arXiv:astro-ph/9709112 [astro-ph]} \BibitemShut {NoStop}%
\bibitem [{\citenamefont {Ludlow}\ \emph {et~al.}(2016)\citenamefont {Ludlow}, \citenamefont {Bose}, \citenamefont {Angulo}, \citenamefont {Wang}, \citenamefont {Hellwing}, \citenamefont {Navarro}, \citenamefont {Cole},\ and\ \citenamefont {Frenk}}]{Ludlow:2016ifl}%
  \BibitemOpen
  \bibfield  {author} {\bibinfo {author} {\bibfnamefont {A.~D.}\ \bibnamefont {Ludlow}}, \bibinfo {author} {\bibfnamefont {S.}~\bibnamefont {Bose}}, \bibinfo {author} {\bibfnamefont {R.~E.}\ \bibnamefont {Angulo}}, \bibinfo {author} {\bibfnamefont {L.}~\bibnamefont {Wang}}, \bibinfo {author} {\bibfnamefont {W.~A.}\ \bibnamefont {Hellwing}}, \bibinfo {author} {\bibfnamefont {J.~F.}\ \bibnamefont {Navarro}}, \bibinfo {author} {\bibfnamefont {S.}~\bibnamefont {Cole}},\ and\ \bibinfo {author} {\bibfnamefont {C.~S.}\ \bibnamefont {Frenk}},\ }\bibfield  {title} {\bibinfo {title} {{The mass\textendash{}concentration\textendash{}redshift relation of cold and warm dark matter haloes}},\ }\href {https://doi.org/10.1093/mnras/stw1046} {\bibfield  {journal} {\bibinfo  {journal} {Mon. Not. Roy. Astron. Soc.}\ }\textbf {\bibinfo {volume} {460}},\ \bibinfo {pages} {1214} (\bibinfo {year} {2016})},\ \Eprint {https://arxiv.org/abs/1601.02624} {arXiv:1601.02624 [astro-ph.CO]} \BibitemShut {NoStop}%
\bibitem [{\citenamefont {{Brown}}\ \emph {et~al.}(2020)\citenamefont {{Brown}}, \citenamefont {{McCarthy}}, \citenamefont {{Diemer}}, \citenamefont {{Font}}, \citenamefont {{Stafford}},\ and\ \citenamefont {{Pfeifer}}}]{Brown:2020lxj}%
  \BibitemOpen
  \bibfield  {author} {\bibinfo {author} {\bibfnamefont {S.~T.}\ \bibnamefont {{Brown}}}, \bibinfo {author} {\bibfnamefont {I.~G.}\ \bibnamefont {{McCarthy}}}, \bibinfo {author} {\bibfnamefont {B.}~\bibnamefont {{Diemer}}}, \bibinfo {author} {\bibfnamefont {A.~S.}\ \bibnamefont {{Font}}}, \bibinfo {author} {\bibfnamefont {S.~G.}\ \bibnamefont {{Stafford}}},\ and\ \bibinfo {author} {\bibfnamefont {S.}~\bibnamefont {{Pfeifer}}},\ }\bibfield  {title} {\bibinfo {title} {{Connecting the structure of dark matter haloes to the primordial power spectrum}},\ }\href {https://doi.org/10.1093/mnras/staa1491} {\bibfield  {journal} {\bibinfo  {journal} {\mnras}\ }\textbf {\bibinfo {volume} {495}},\ \bibinfo {pages} {4994} (\bibinfo {year} {2020})},\ \Eprint {https://arxiv.org/abs/2005.12933} {arXiv:2005.12933 [astro-ph.CO]} \BibitemShut {NoStop}%
\bibitem [{\citenamefont {Griffen}\ \emph {et~al.}(2016)\citenamefont {Griffen}, \citenamefont {Ji}, \citenamefont {Dooley}, \citenamefont {Gómez}, \citenamefont {Vogelsberger}, \citenamefont {O’Shea},\ and\ \citenamefont {Frebel}}]{Griffen_2016}%
  \BibitemOpen
  \bibfield  {author} {\bibinfo {author} {\bibfnamefont {B.~F.}\ \bibnamefont {Griffen}}, \bibinfo {author} {\bibfnamefont {A.~P.}\ \bibnamefont {Ji}}, \bibinfo {author} {\bibfnamefont {G.~A.}\ \bibnamefont {Dooley}}, \bibinfo {author} {\bibfnamefont {F.~A.}\ \bibnamefont {Gómez}}, \bibinfo {author} {\bibfnamefont {M.}~\bibnamefont {Vogelsberger}}, \bibinfo {author} {\bibfnamefont {B.~W.}\ \bibnamefont {O’Shea}},\ and\ \bibinfo {author} {\bibfnamefont {A.}~\bibnamefont {Frebel}},\ }\bibfield  {title} {\bibinfo {title} {The caterpillar project: A large suite of milky way sized halos},\ }\href {https://doi.org/10.3847/0004-637x/818/1/10} {\bibfield  {journal} {\bibinfo  {journal} {The Astrophysical Journal}\ }\textbf {\bibinfo {volume} {818}},\ \bibinfo {pages} {10} (\bibinfo {year} {2016})}\BibitemShut {NoStop}%
\bibitem [{\citenamefont {Conroy}\ \emph {et~al.}(2009)\citenamefont {Conroy}, \citenamefont {Gunn},\ and\ \citenamefont {White}}]{Conroy:2008mp}%
  \BibitemOpen
  \bibfield  {author} {\bibinfo {author} {\bibfnamefont {C.}~\bibnamefont {Conroy}}, \bibinfo {author} {\bibfnamefont {J.~E.}\ \bibnamefont {Gunn}},\ and\ \bibinfo {author} {\bibfnamefont {M.}~\bibnamefont {White}},\ }\bibfield  {title} {\bibinfo {title} {{The propagation of uncertainties in stellar population synthesis modeling I: The relevance of uncertain aspects of stellar evolution and the IMF to the derived physical properties of galaxies}},\ }\href {https://doi.org/10.1088/0004-637X/699/1/486} {\bibfield  {journal} {\bibinfo  {journal} {Astrophys. J.}\ }\textbf {\bibinfo {volume} {699}},\ \bibinfo {pages} {486} (\bibinfo {year} {2009})},\ \Eprint {https://arxiv.org/abs/0809.4261} {arXiv:0809.4261 [astro-ph]} \BibitemShut {NoStop}%
\bibitem [{\citenamefont {Conroy}\ and\ \citenamefont {Gunn}(2010)}]{Conroy:2009vq}%
  \BibitemOpen
  \bibfield  {author} {\bibinfo {author} {\bibfnamefont {C.}~\bibnamefont {Conroy}}\ and\ \bibinfo {author} {\bibfnamefont {J.~E.}\ \bibnamefont {Gunn}},\ }\bibfield  {title} {\bibinfo {title} {{The propagation of uncertainties in stellar population synthesis modeling III: model calibration, comparison, and evaluation}},\ }\href {https://doi.org/10.1088/0004-637X/712/2/833} {\bibfield  {journal} {\bibinfo  {journal} {Astrophys. J.}\ }\textbf {\bibinfo {volume} {712}},\ \bibinfo {pages} {833} (\bibinfo {year} {2010})},\ \Eprint {https://arxiv.org/abs/0911.3151} {arXiv:0911.3151 [astro-ph.CO]} \BibitemShut {NoStop}%
\bibitem [{\citenamefont {Navarro}\ \emph {et~al.}(1997)\citenamefont {Navarro}, \citenamefont {Frenk},\ and\ \citenamefont {White}}]{Navarro:1996gj}%
  \BibitemOpen
  \bibfield  {author} {\bibinfo {author} {\bibfnamefont {J.~F.}\ \bibnamefont {Navarro}}, \bibinfo {author} {\bibfnamefont {C.~S.}\ \bibnamefont {Frenk}},\ and\ \bibinfo {author} {\bibfnamefont {S.~D.~M.}\ \bibnamefont {White}},\ }\bibfield  {title} {\bibinfo {title} {{A Universal density profile from hierarchical clustering}},\ }\href {https://doi.org/10.1086/304888} {\bibfield  {journal} {\bibinfo  {journal} {Astrophys. J.}\ }\textbf {\bibinfo {volume} {490}},\ \bibinfo {pages} {493} (\bibinfo {year} {1997})},\ \Eprint {https://arxiv.org/abs/astro-ph/9611107} {arXiv:astro-ph/9611107} \BibitemShut {NoStop}%
\bibitem [{\citenamefont {Scott}(1992)}]{scott1992}%
  \BibitemOpen
  \bibfield  {author} {\bibinfo {author} {\bibfnamefont {D.~W.}\ \bibnamefont {Scott}},\ }\href {https://doi.org/10.1002/9780470316849} {\emph {\bibinfo {title} {Multivariate Density Estimation: Theory, Practice, and Visualization}}}\ (\bibinfo  {publisher} {Wiley},\ \bibinfo {year} {1992})\BibitemShut {NoStop}%
\bibitem [{\citenamefont {Pedregosa}\ \emph {et~al.}(2011)\citenamefont {Pedregosa}, \citenamefont {Varoquaux}, \citenamefont {Gramfort}, \citenamefont {Michel}, \citenamefont {Thirion}, \citenamefont {Grisel}, \citenamefont {Blondel}, \citenamefont {Prettenhofer}, \citenamefont {Weiss}, \citenamefont {Dubourg}, \citenamefont {Vanderplas}, \citenamefont {Passos}, \citenamefont {Cournapeau}, \citenamefont {Brucher}, \citenamefont {Perrot},\ and\ \citenamefont {Duchesnay}}]{scikit-learn}%
  \BibitemOpen
  \bibfield  {author} {\bibinfo {author} {\bibfnamefont {F.}~\bibnamefont {Pedregosa}}, \bibinfo {author} {\bibfnamefont {G.}~\bibnamefont {Varoquaux}}, \bibinfo {author} {\bibfnamefont {A.}~\bibnamefont {Gramfort}}, \bibinfo {author} {\bibfnamefont {V.}~\bibnamefont {Michel}}, \bibinfo {author} {\bibfnamefont {B.}~\bibnamefont {Thirion}}, \bibinfo {author} {\bibfnamefont {O.}~\bibnamefont {Grisel}}, \bibinfo {author} {\bibfnamefont {M.}~\bibnamefont {Blondel}}, \bibinfo {author} {\bibfnamefont {P.}~\bibnamefont {Prettenhofer}}, \bibinfo {author} {\bibfnamefont {R.}~\bibnamefont {Weiss}}, \bibinfo {author} {\bibfnamefont {V.}~\bibnamefont {Dubourg}}, \bibinfo {author} {\bibfnamefont {J.}~\bibnamefont {Vanderplas}}, \bibinfo {author} {\bibfnamefont {A.}~\bibnamefont {Passos}}, \bibinfo {author} {\bibfnamefont {D.}~\bibnamefont {Cournapeau}}, \bibinfo {author} {\bibfnamefont {M.}~\bibnamefont {Brucher}}, \bibinfo {author} {\bibfnamefont {M.}~\bibnamefont {Perrot}},\ and\ \bibinfo {author} {\bibfnamefont
  {E.}~\bibnamefont {Duchesnay}},\ }\bibfield  {title} {\bibinfo {title} {Scikit-learn: Machine learning in {P}ython},\ }\href@noop {} {\bibfield  {journal} {\bibinfo  {journal} {Journal of Machine Learning Research}\ }\textbf {\bibinfo {volume} {12}},\ \bibinfo {pages} {2825} (\bibinfo {year} {2011})}\BibitemShut {NoStop}%
\bibitem [{\citenamefont {Manwadkar}\ and\ \citenamefont {Kravtsov}(2022)}]{Manwadkar_2022}%
  \BibitemOpen
  \bibfield  {author} {\bibinfo {author} {\bibfnamefont {V.}~\bibnamefont {Manwadkar}}\ and\ \bibinfo {author} {\bibfnamefont {A.~V.}\ \bibnamefont {Kravtsov}},\ }\bibfield  {title} {\bibinfo {title} {Forward-modelling the luminosity, distance, and size distributions of the milky way satellites},\ }\href {https://doi.org/10.1093/mnras/stac2452} {\bibfield  {journal} {\bibinfo  {journal} {Monthly Notices of the Royal Astronomical Society}\ }\textbf {\bibinfo {volume} {516}},\ \bibinfo {pages} {3944–3971} (\bibinfo {year} {2022})}\BibitemShut {NoStop}%
\bibitem [{\citenamefont {Wilks}(1938)}]{Wilks:1938dza}%
  \BibitemOpen
  \bibfield  {author} {\bibinfo {author} {\bibfnamefont {S.~S.}\ \bibnamefont {Wilks}},\ }\bibfield  {title} {\bibinfo {title} {{The Large-Sample Distribution of the Likelihood Ratio for Testing Composite Hypotheses}},\ }\href {https://doi.org/10.1214/aoms/1177732360} {\bibfield  {journal} {\bibinfo  {journal} {Annals Math. Statist.}\ }\textbf {\bibinfo {volume} {9}},\ \bibinfo {pages} {60} (\bibinfo {year} {1938})}\BibitemShut {NoStop}%
\bibitem [{\citenamefont {Bagley}\ \emph {et~al.}(2023)\citenamefont {Bagley}, \citenamefont {Finkelstein}, \citenamefont {Koekemoer}, \citenamefont {Ferguson}, \citenamefont {Arrabal~Haro}, \citenamefont {Dickinson}, \citenamefont {Kartaltepe}, \citenamefont {Papovich}, \citenamefont {Pérez-González}, \citenamefont {Pirzkal}, \citenamefont {Somerville}, \citenamefont {Willmer}, \citenamefont {Yang}, \citenamefont {Yung}, \citenamefont {Fontana}, \citenamefont {Grazian}, \citenamefont {Grogin}, \citenamefont {Hirschmann}, \citenamefont {Kewley}, \citenamefont {Kirkpatrick}, \citenamefont {Kocevski}, \citenamefont {Lotz}, \citenamefont {Medrano}, \citenamefont {Morales}, \citenamefont {Pentericci}, \citenamefont {Ravindranath}, \citenamefont {Trump}, \citenamefont {Wilkins}, \citenamefont {Calabrò}, \citenamefont {Cooper}, \citenamefont {Costantin}, \citenamefont {de~la Vega}, \citenamefont {Hilbert}, \citenamefont {Hutchison}, \citenamefont {Larson}, \citenamefont {Lucas}, \citenamefont {McGrath},
  \citenamefont {Ryan}, \citenamefont {Wang},\ and\ \citenamefont {Wuyts}}]{Bagley_2023}%
  \BibitemOpen
  \bibfield  {author} {\bibinfo {author} {\bibfnamefont {M.~B.}\ \bibnamefont {Bagley}}, \bibinfo {author} {\bibfnamefont {S.~L.}\ \bibnamefont {Finkelstein}}, \bibinfo {author} {\bibfnamefont {A.~M.}\ \bibnamefont {Koekemoer}}, \bibinfo {author} {\bibfnamefont {H.~C.}\ \bibnamefont {Ferguson}}, \bibinfo {author} {\bibfnamefont {P.}~\bibnamefont {Arrabal~Haro}}, \bibinfo {author} {\bibfnamefont {M.}~\bibnamefont {Dickinson}}, \bibinfo {author} {\bibfnamefont {J.~S.}\ \bibnamefont {Kartaltepe}}, \bibinfo {author} {\bibfnamefont {C.}~\bibnamefont {Papovich}}, \bibinfo {author} {\bibfnamefont {P.~G.}\ \bibnamefont {Pérez-González}}, \bibinfo {author} {\bibfnamefont {N.}~\bibnamefont {Pirzkal}}, \bibinfo {author} {\bibfnamefont {R.~S.}\ \bibnamefont {Somerville}}, \bibinfo {author} {\bibfnamefont {C.~N.~A.}\ \bibnamefont {Willmer}}, \bibinfo {author} {\bibfnamefont {G.}~\bibnamefont {Yang}}, \bibinfo {author} {\bibfnamefont {L.~Y.~A.}\ \bibnamefont {Yung}}, \bibinfo {author} {\bibfnamefont {A.}~\bibnamefont
  {Fontana}}, \bibinfo {author} {\bibfnamefont {A.}~\bibnamefont {Grazian}}, \bibinfo {author} {\bibfnamefont {N.~A.}\ \bibnamefont {Grogin}}, \bibinfo {author} {\bibfnamefont {M.}~\bibnamefont {Hirschmann}}, \bibinfo {author} {\bibfnamefont {L.~J.}\ \bibnamefont {Kewley}}, \bibinfo {author} {\bibfnamefont {A.}~\bibnamefont {Kirkpatrick}}, \bibinfo {author} {\bibfnamefont {D.~D.}\ \bibnamefont {Kocevski}}, \bibinfo {author} {\bibfnamefont {J.~M.}\ \bibnamefont {Lotz}}, \bibinfo {author} {\bibfnamefont {A.}~\bibnamefont {Medrano}}, \bibinfo {author} {\bibfnamefont {A.~M.}\ \bibnamefont {Morales}}, \bibinfo {author} {\bibfnamefont {L.}~\bibnamefont {Pentericci}}, \bibinfo {author} {\bibfnamefont {S.}~\bibnamefont {Ravindranath}}, \bibinfo {author} {\bibfnamefont {J.~R.}\ \bibnamefont {Trump}}, \bibinfo {author} {\bibfnamefont {S.~M.}\ \bibnamefont {Wilkins}}, \bibinfo {author} {\bibfnamefont {A.}~\bibnamefont {Calabrò}}, \bibinfo {author} {\bibfnamefont {M.~C.}\ \bibnamefont {Cooper}}, \bibinfo {author}
  {\bibfnamefont {L.}~\bibnamefont {Costantin}}, \bibinfo {author} {\bibfnamefont {A.}~\bibnamefont {de~la Vega}}, \bibinfo {author} {\bibfnamefont {B.}~\bibnamefont {Hilbert}}, \bibinfo {author} {\bibfnamefont {T.~A.}\ \bibnamefont {Hutchison}}, \bibinfo {author} {\bibfnamefont {R.~L.}\ \bibnamefont {Larson}}, \bibinfo {author} {\bibfnamefont {R.~A.}\ \bibnamefont {Lucas}}, \bibinfo {author} {\bibfnamefont {E.~J.}\ \bibnamefont {McGrath}}, \bibinfo {author} {\bibfnamefont {R.}~\bibnamefont {Ryan}}, \bibinfo {author} {\bibfnamefont {X.}~\bibnamefont {Wang}},\ and\ \bibinfo {author} {\bibfnamefont {S.}~\bibnamefont {Wuyts}},\ }\bibfield  {title} {\bibinfo {title} {Ceers epoch 1 nircam imaging: Reduction methods and simulations enabling early jwst science results},\ }\href {https://doi.org/10.3847/2041-8213/acbb08} {\bibfield  {journal} {\bibinfo  {journal} {The Astrophysical Journal Letters}\ }\textbf {\bibinfo {volume} {946}},\ \bibinfo {pages} {L12} (\bibinfo {year} {2023})}\BibitemShut {NoStop}%
\bibitem [{\citenamefont {{Labb{\'e}}}\ \emph {et~al.}(2023)\citenamefont {{Labb{\'e}}}, \citenamefont {{van Dokkum}}, \citenamefont {{Nelson}}, \citenamefont {{Bezanson}}, \citenamefont {{Suess}}, \citenamefont {{Leja}}, \citenamefont {{Brammer}}, \citenamefont {{Whitaker}}, \citenamefont {{Mathews}}, \citenamefont {{Stefanon}},\ and\ \citenamefont {{Wang}}}]{2023Natur.616..266L}%
  \BibitemOpen
  \bibfield  {author} {\bibinfo {author} {\bibfnamefont {I.}~\bibnamefont {{Labb{\'e}}}}, \bibinfo {author} {\bibfnamefont {P.}~\bibnamefont {{van Dokkum}}}, \bibinfo {author} {\bibfnamefont {E.}~\bibnamefont {{Nelson}}}, \bibinfo {author} {\bibfnamefont {R.}~\bibnamefont {{Bezanson}}}, \bibinfo {author} {\bibfnamefont {K.~A.}\ \bibnamefont {{Suess}}}, \bibinfo {author} {\bibfnamefont {J.}~\bibnamefont {{Leja}}}, \bibinfo {author} {\bibfnamefont {G.}~\bibnamefont {{Brammer}}}, \bibinfo {author} {\bibfnamefont {K.}~\bibnamefont {{Whitaker}}}, \bibinfo {author} {\bibfnamefont {E.}~\bibnamefont {{Mathews}}}, \bibinfo {author} {\bibfnamefont {M.}~\bibnamefont {{Stefanon}}},\ and\ \bibinfo {author} {\bibfnamefont {B.}~\bibnamefont {{Wang}}},\ }\bibfield  {title} {\bibinfo {title} {{A population of red candidate massive galaxies 600 Myr after the Big Bang}},\ }\href {https://doi.org/10.1038/s41586-023-05786-2} {\bibfield  {journal} {\bibinfo  {journal} {\nat}\ }\textbf {\bibinfo {volume} {616}},\ \bibinfo {pages}
  {266} (\bibinfo {year} {2023})},\ \Eprint {https://arxiv.org/abs/2207.12446} {arXiv:2207.12446 [astro-ph.GA]} \BibitemShut {NoStop}%
\bibitem [{\citenamefont {Chworowsky}\ \emph {et~al.}(2023)\citenamefont {Chworowsky}, \citenamefont {Finkelstein}, \citenamefont {Boylan-Kolchin}, \citenamefont {McGrath}, \citenamefont {Iyer}, \citenamefont {Papovich}, \citenamefont {Dickinson}, \citenamefont {Taylor}, \citenamefont {Yung}, \citenamefont {Haro}, \citenamefont {Bagley}, \citenamefont {Backhaus}, \citenamefont {Bhatawdekar}, \citenamefont {Cheng}, \citenamefont {Cleri}, \citenamefont {Cole}, \citenamefont {Cooper}, \citenamefont {Costantin}, \citenamefont {Dekel}, \citenamefont {Franco}, \citenamefont {Fujimoto}, \citenamefont {Hayward}, \citenamefont {Holwerda}, \citenamefont {Huertas-Company}, \citenamefont {Hirschmann}, \citenamefont {Hutchison}, \citenamefont {Koekemoer}, \citenamefont {Larson}, \citenamefont {Li}, \citenamefont {Long}, \citenamefont {Lucas}, \citenamefont {Pirzkal}, \citenamefont {Rodighiero}, \citenamefont {Somerville}, \citenamefont {Vanderhoof}, \citenamefont {de~la Vega}, \citenamefont {Wilkins}, \citenamefont
  {Yang},\ and\ \citenamefont {Zavala}}]{chworowsky2023evidence}%
  \BibitemOpen
  \bibfield  {author} {\bibinfo {author} {\bibfnamefont {K.}~\bibnamefont {Chworowsky}}, \bibinfo {author} {\bibfnamefont {S.~L.}\ \bibnamefont {Finkelstein}}, \bibinfo {author} {\bibfnamefont {M.}~\bibnamefont {Boylan-Kolchin}}, \bibinfo {author} {\bibfnamefont {E.~J.}\ \bibnamefont {McGrath}}, \bibinfo {author} {\bibfnamefont {K.~G.}\ \bibnamefont {Iyer}}, \bibinfo {author} {\bibfnamefont {C.}~\bibnamefont {Papovich}}, \bibinfo {author} {\bibfnamefont {M.}~\bibnamefont {Dickinson}}, \bibinfo {author} {\bibfnamefont {A.~J.}\ \bibnamefont {Taylor}}, \bibinfo {author} {\bibfnamefont {L.~Y.~A.}\ \bibnamefont {Yung}}, \bibinfo {author} {\bibfnamefont {P.~A.}\ \bibnamefont {Haro}}, \bibinfo {author} {\bibfnamefont {M.~B.}\ \bibnamefont {Bagley}}, \bibinfo {author} {\bibfnamefont {B.~E.}\ \bibnamefont {Backhaus}}, \bibinfo {author} {\bibfnamefont {R.}~\bibnamefont {Bhatawdekar}}, \bibinfo {author} {\bibfnamefont {Y.}~\bibnamefont {Cheng}}, \bibinfo {author} {\bibfnamefont {N.~J.}\ \bibnamefont {Cleri}}, \bibinfo
  {author} {\bibfnamefont {J.~W.}\ \bibnamefont {Cole}}, \bibinfo {author} {\bibfnamefont {M.~C.}\ \bibnamefont {Cooper}}, \bibinfo {author} {\bibfnamefont {L.}~\bibnamefont {Costantin}}, \bibinfo {author} {\bibfnamefont {A.}~\bibnamefont {Dekel}}, \bibinfo {author} {\bibfnamefont {M.}~\bibnamefont {Franco}}, \bibinfo {author} {\bibfnamefont {S.}~\bibnamefont {Fujimoto}}, \bibinfo {author} {\bibfnamefont {C.~C.}\ \bibnamefont {Hayward}}, \bibinfo {author} {\bibfnamefont {B.~W.}\ \bibnamefont {Holwerda}}, \bibinfo {author} {\bibfnamefont {M.}~\bibnamefont {Huertas-Company}}, \bibinfo {author} {\bibfnamefont {M.}~\bibnamefont {Hirschmann}}, \bibinfo {author} {\bibfnamefont {T.~A.}\ \bibnamefont {Hutchison}}, \bibinfo {author} {\bibfnamefont {A.~M.}\ \bibnamefont {Koekemoer}}, \bibinfo {author} {\bibfnamefont {R.~L.}\ \bibnamefont {Larson}}, \bibinfo {author} {\bibfnamefont {Z.}~\bibnamefont {Li}}, \bibinfo {author} {\bibfnamefont {A.~S.}\ \bibnamefont {Long}}, \bibinfo {author} {\bibfnamefont {R.~A.}\
  \bibnamefont {Lucas}}, \bibinfo {author} {\bibfnamefont {N.}~\bibnamefont {Pirzkal}}, \bibinfo {author} {\bibfnamefont {G.}~\bibnamefont {Rodighiero}}, \bibinfo {author} {\bibfnamefont {R.~S.}\ \bibnamefont {Somerville}}, \bibinfo {author} {\bibfnamefont {B.~N.}\ \bibnamefont {Vanderhoof}}, \bibinfo {author} {\bibfnamefont {A.}~\bibnamefont {de~la Vega}}, \bibinfo {author} {\bibfnamefont {S.~M.}\ \bibnamefont {Wilkins}}, \bibinfo {author} {\bibfnamefont {G.}~\bibnamefont {Yang}},\ and\ \bibinfo {author} {\bibfnamefont {J.~A.}\ \bibnamefont {Zavala}},\ }\href@noop {} {\bibinfo {title} {Evidence for a shallow evolution in the volume densities of massive galaxies at $z=4$ to $8$ from ceers}} (\bibinfo {year} {2023}),\ \Eprint {https://arxiv.org/abs/2311.14804} {arXiv:2311.14804 [astro-ph.GA]} \BibitemShut {NoStop}%
\bibitem [{\citenamefont {Sun}\ \emph {et~al.}(2023)\citenamefont {Sun}, \citenamefont {Faucher-Giguère}, \citenamefont {Hayward}, \citenamefont {Shen}, \citenamefont {Wetzel},\ and\ \citenamefont {Cochrane}}]{sun2023bursty}%
  \BibitemOpen
  \bibfield  {author} {\bibinfo {author} {\bibfnamefont {G.}~\bibnamefont {Sun}}, \bibinfo {author} {\bibfnamefont {C.-A.}\ \bibnamefont {Faucher-Giguère}}, \bibinfo {author} {\bibfnamefont {C.~C.}\ \bibnamefont {Hayward}}, \bibinfo {author} {\bibfnamefont {X.}~\bibnamefont {Shen}}, \bibinfo {author} {\bibfnamefont {A.}~\bibnamefont {Wetzel}},\ and\ \bibinfo {author} {\bibfnamefont {R.~K.}\ \bibnamefont {Cochrane}},\ }\href@noop {} {\bibinfo {title} {Bursty star formation naturally explains the abundance of bright galaxies at cosmic dawn}} (\bibinfo {year} {2023}),\ \Eprint {https://arxiv.org/abs/2307.15305} {arXiv:2307.15305 [astro-ph.GA]} \BibitemShut {NoStop}%
\bibitem [{\citenamefont {Ir\v{s}i\v{c}}\ \emph {et~al.}(2024)\citenamefont {Ir\v{s}i\v{c}} \emph {et~al.}}]{Irsic:2023equ}%
  \BibitemOpen
  \bibfield  {author} {\bibinfo {author} {\bibfnamefont {V.}~\bibnamefont {Ir\v{s}i\v{c}}} \emph {et~al.},\ }\bibfield  {title} {\bibinfo {title} {{Unveiling dark matter free streaming at the smallest scales with the high redshift Lyman-alpha forest}},\ }\href {https://doi.org/10.1103/PhysRevD.109.043511} {\bibfield  {journal} {\bibinfo  {journal} {Phys. Rev. D}\ }\textbf {\bibinfo {volume} {109}},\ \bibinfo {pages} {043511} (\bibinfo {year} {2024})},\ \Eprint {https://arxiv.org/abs/2309.04533} {arXiv:2309.04533 [astro-ph.CO]} \BibitemShut {NoStop}%
\bibitem [{\citenamefont {Lentati}\ \emph {et~al.}(2015)\citenamefont {Lentati} \emph {et~al.}}]{EPTA:2015qep}%
  \BibitemOpen
  \bibfield  {author} {\bibinfo {author} {\bibfnamefont {L.}~\bibnamefont {Lentati}} \emph {et~al.} (\bibinfo {collaboration} {EPTA}),\ }\bibfield  {title} {\bibinfo {title} {{European Pulsar Timing Array Limits On An Isotropic Stochastic Gravitational-Wave Background}},\ }\href {https://doi.org/10.1093/mnras/stv1538} {\bibfield  {journal} {\bibinfo  {journal} {Mon. Not. Roy. Astron. Soc.}\ }\textbf {\bibinfo {volume} {453}},\ \bibinfo {pages} {2576} (\bibinfo {year} {2015})},\ \Eprint {https://arxiv.org/abs/1504.03692} {arXiv:1504.03692 [astro-ph.CO]} \BibitemShut {NoStop}%
\bibitem [{\citenamefont {Inomata}\ and\ \citenamefont {Nakama}(2019)}]{Inomata:2018epa}%
  \BibitemOpen
  \bibfield  {author} {\bibinfo {author} {\bibfnamefont {K.}~\bibnamefont {Inomata}}\ and\ \bibinfo {author} {\bibfnamefont {T.}~\bibnamefont {Nakama}},\ }\bibfield  {title} {\bibinfo {title} {{Gravitational waves induced by scalar perturbations as probes of the small-scale primordial spectrum}},\ }\href {https://doi.org/10.1103/PhysRevD.99.043511} {\bibfield  {journal} {\bibinfo  {journal} {Phys. Rev. D}\ }\textbf {\bibinfo {volume} {99}},\ \bibinfo {pages} {043511} (\bibinfo {year} {2019})},\ \Eprint {https://arxiv.org/abs/1812.00674} {arXiv:1812.00674 [astro-ph.CO]} \BibitemShut {NoStop}%
\bibitem [{\citenamefont {Byrnes}\ \emph {et~al.}(2019)\citenamefont {Byrnes}, \citenamefont {Cole},\ and\ \citenamefont {Patil}}]{Byrnes:2018txb}%
  \BibitemOpen
  \bibfield  {author} {\bibinfo {author} {\bibfnamefont {C.~T.}\ \bibnamefont {Byrnes}}, \bibinfo {author} {\bibfnamefont {P.~S.}\ \bibnamefont {Cole}},\ and\ \bibinfo {author} {\bibfnamefont {S.~P.}\ \bibnamefont {Patil}},\ }\bibfield  {title} {\bibinfo {title} {{Steepest growth of the power spectrum and primordial black holes}},\ }\href {https://doi.org/10.1088/1475-7516/2019/06/028} {\bibfield  {journal} {\bibinfo  {journal} {JCAP}\ }\textbf {\bibinfo {volume} {06}},\ \bibinfo {pages} {028}},\ \Eprint {https://arxiv.org/abs/1811.11158} {arXiv:1811.11158 [astro-ph.CO]} \BibitemShut {NoStop}%
\bibitem [{\citenamefont {Green}\ and\ \citenamefont {Kavanagh}(2021)}]{Green:2020jor}%
  \BibitemOpen
  \bibfield  {author} {\bibinfo {author} {\bibfnamefont {A.~M.}\ \bibnamefont {Green}}\ and\ \bibinfo {author} {\bibfnamefont {B.~J.}\ \bibnamefont {Kavanagh}},\ }\bibfield  {title} {\bibinfo {title} {{Primordial Black Holes as a dark matter candidate}},\ }\href {https://doi.org/10.1088/1361-6471/abc534} {\bibfield  {journal} {\bibinfo  {journal} {J. Phys. G}\ }\textbf {\bibinfo {volume} {48}},\ \bibinfo {pages} {043001} (\bibinfo {year} {2021})},\ \Eprint {https://arxiv.org/abs/2007.10722} {arXiv:2007.10722 [astro-ph.CO]} \BibitemShut {NoStop}%
\bibitem [{\citenamefont {Carr}\ \emph {et~al.}(2021)\citenamefont {Carr}, \citenamefont {Kohri}, \citenamefont {Sendouda},\ and\ \citenamefont {Yokoyama}}]{Carr:2020gox}%
  \BibitemOpen
  \bibfield  {author} {\bibinfo {author} {\bibfnamefont {B.}~\bibnamefont {Carr}}, \bibinfo {author} {\bibfnamefont {K.}~\bibnamefont {Kohri}}, \bibinfo {author} {\bibfnamefont {Y.}~\bibnamefont {Sendouda}},\ and\ \bibinfo {author} {\bibfnamefont {J.}~\bibnamefont {Yokoyama}},\ }\bibfield  {title} {\bibinfo {title} {{Constraints on primordial black holes}},\ }\href {https://doi.org/10.1088/1361-6633/ac1e31} {\bibfield  {journal} {\bibinfo  {journal} {Rept. Prog. Phys.}\ }\textbf {\bibinfo {volume} {84}},\ \bibinfo {pages} {116902} (\bibinfo {year} {2021})},\ \Eprint {https://arxiv.org/abs/2002.12778} {arXiv:2002.12778 [astro-ph.CO]} \BibitemShut {NoStop}%
\bibitem [{\citenamefont {Kawasaki}\ \emph {et~al.}(2016)\citenamefont {Kawasaki}, \citenamefont {Kusenko}, \citenamefont {Tada},\ and\ \citenamefont {Yanagida}}]{Kawasaki:2016pql}%
  \BibitemOpen
  \bibfield  {author} {\bibinfo {author} {\bibfnamefont {M.}~\bibnamefont {Kawasaki}}, \bibinfo {author} {\bibfnamefont {A.}~\bibnamefont {Kusenko}}, \bibinfo {author} {\bibfnamefont {Y.}~\bibnamefont {Tada}},\ and\ \bibinfo {author} {\bibfnamefont {T.~T.}\ \bibnamefont {Yanagida}},\ }\bibfield  {title} {\bibinfo {title} {{Primordial black holes as dark matter in supergravity inflation models}},\ }\href {https://doi.org/10.1103/PhysRevD.94.083523} {\bibfield  {journal} {\bibinfo  {journal} {Phys. Rev. D}\ }\textbf {\bibinfo {volume} {94}},\ \bibinfo {pages} {083523} (\bibinfo {year} {2016})},\ \Eprint {https://arxiv.org/abs/1606.07631} {arXiv:1606.07631 [astro-ph.CO]} \BibitemShut {NoStop}%
\bibitem [{\citenamefont {Chluba}\ \emph {et~al.}(2012)\citenamefont {Chluba}, \citenamefont {Erickcek},\ and\ \citenamefont {Ben-Dayan}}]{Chluba:2012we}%
  \BibitemOpen
  \bibfield  {author} {\bibinfo {author} {\bibfnamefont {J.}~\bibnamefont {Chluba}}, \bibinfo {author} {\bibfnamefont {A.~L.}\ \bibnamefont {Erickcek}},\ and\ \bibinfo {author} {\bibfnamefont {I.}~\bibnamefont {Ben-Dayan}},\ }\bibfield  {title} {\bibinfo {title} {{Probing the inflaton: Small-scale power spectrum constraints from measurements of the CMB energy spectrum}},\ }\href {https://doi.org/10.1088/0004-637X/758/2/76} {\bibfield  {journal} {\bibinfo  {journal} {Astrophys. J.}\ }\textbf {\bibinfo {volume} {758}},\ \bibinfo {pages} {76} (\bibinfo {year} {2012})},\ \Eprint {https://arxiv.org/abs/1203.2681} {arXiv:1203.2681 [astro-ph.CO]} \BibitemShut {NoStop}%
\bibitem [{\citenamefont {Mather}\ \emph {et~al.}(1994)\citenamefont {Mather} \emph {et~al.}}]{Mather:1993ij}%
  \BibitemOpen
  \bibfield  {author} {\bibinfo {author} {\bibfnamefont {J.~C.}\ \bibnamefont {Mather}} \emph {et~al.},\ }\bibfield  {title} {\bibinfo {title} {{Measurement of the Cosmic Microwave Background spectrum by the COBE FIRAS instrument}},\ }\href {https://doi.org/10.1086/173574} {\bibfield  {journal} {\bibinfo  {journal} {Astrophys. J.}\ }\textbf {\bibinfo {volume} {420}},\ \bibinfo {pages} {439} (\bibinfo {year} {1994})}\BibitemShut {NoStop}%
\bibitem [{\citenamefont {Fixsen}\ \emph {et~al.}(1996)\citenamefont {Fixsen}, \citenamefont {Cheng}, \citenamefont {Gales}, \citenamefont {Mather}, \citenamefont {Shafer},\ and\ \citenamefont {Wright}}]{Fixsen:1996nj}%
  \BibitemOpen
  \bibfield  {author} {\bibinfo {author} {\bibfnamefont {D.~J.}\ \bibnamefont {Fixsen}}, \bibinfo {author} {\bibfnamefont {E.~S.}\ \bibnamefont {Cheng}}, \bibinfo {author} {\bibfnamefont {J.~M.}\ \bibnamefont {Gales}}, \bibinfo {author} {\bibfnamefont {J.~C.}\ \bibnamefont {Mather}}, \bibinfo {author} {\bibfnamefont {R.~A.}\ \bibnamefont {Shafer}},\ and\ \bibinfo {author} {\bibfnamefont {E.~L.}\ \bibnamefont {Wright}},\ }\bibfield  {title} {\bibinfo {title} {{The Cosmic Microwave Background spectrum from the full COBE FIRAS data set}},\ }\href {https://doi.org/10.1086/178173} {\bibfield  {journal} {\bibinfo  {journal} {Astrophys. J.}\ }\textbf {\bibinfo {volume} {473}},\ \bibinfo {pages} {576} (\bibinfo {year} {1996})},\ \Eprint {https://arxiv.org/abs/astro-ph/9605054} {arXiv:astro-ph/9605054} \BibitemShut {NoStop}%
\bibitem [{\citenamefont {Cyr}\ \emph {et~al.}(2024)\citenamefont {Cyr}, \citenamefont {Kite}, \citenamefont {Chluba}, \citenamefont {Hill}, \citenamefont {Jeong}, \citenamefont {Acharya}, \citenamefont {Bolliet},\ and\ \citenamefont {Patil}}]{Cyr:2023pgw}%
  \BibitemOpen
  \bibfield  {author} {\bibinfo {author} {\bibfnamefont {B.}~\bibnamefont {Cyr}}, \bibinfo {author} {\bibfnamefont {T.}~\bibnamefont {Kite}}, \bibinfo {author} {\bibfnamefont {J.}~\bibnamefont {Chluba}}, \bibinfo {author} {\bibfnamefont {J.~C.}\ \bibnamefont {Hill}}, \bibinfo {author} {\bibfnamefont {D.}~\bibnamefont {Jeong}}, \bibinfo {author} {\bibfnamefont {S.~K.}\ \bibnamefont {Acharya}}, \bibinfo {author} {\bibfnamefont {B.}~\bibnamefont {Bolliet}},\ and\ \bibinfo {author} {\bibfnamefont {S.~P.}\ \bibnamefont {Patil}},\ }\bibfield  {title} {\bibinfo {title} {{Disentangling the primordial nature of stochastic gravitational wave backgrounds with CMB spectral distortions}},\ }\href {https://doi.org/10.1093/mnras/stad3861} {\bibfield  {journal} {\bibinfo  {journal} {Mon. Not. Roy. Astron. Soc.}\ }\textbf {\bibinfo {volume} {528}},\ \bibinfo {pages} {883} (\bibinfo {year} {2024})},\ \Eprint {https://arxiv.org/abs/2309.02366} {arXiv:2309.02366 [astro-ph.CO]} \BibitemShut {NoStop}%
\bibitem [{\citenamefont {Akrami}\ \emph {et~al.}(2020)\citenamefont {Akrami} \emph {et~al.}}]{Planck:2018jri}%
  \BibitemOpen
  \bibfield  {author} {\bibinfo {author} {\bibfnamefont {Y.}~\bibnamefont {Akrami}} \emph {et~al.} (\bibinfo {collaboration} {Planck}),\ }\bibfield  {title} {\bibinfo {title} {{Planck 2018 results. X. Constraints on inflation}},\ }\href {https://doi.org/10.1051/0004-6361/201833887} {\bibfield  {journal} {\bibinfo  {journal} {Astron. Astrophys.}\ }\textbf {\bibinfo {volume} {641}},\ \bibinfo {pages} {A10} (\bibinfo {year} {2020})},\ \Eprint {https://arxiv.org/abs/1807.06211} {arXiv:1807.06211 [astro-ph.CO]} \BibitemShut {NoStop}%
\bibitem [{\citenamefont {{Bird}}\ \emph {et~al.}(2011)\citenamefont {{Bird}}, \citenamefont {{Peiris}}, \citenamefont {{Viel}},\ and\ \citenamefont {{Verde}}}]{Bird_2011}%
  \BibitemOpen
  \bibfield  {author} {\bibinfo {author} {\bibfnamefont {S.}~\bibnamefont {{Bird}}}, \bibinfo {author} {\bibfnamefont {H.~V.}\ \bibnamefont {{Peiris}}}, \bibinfo {author} {\bibfnamefont {M.}~\bibnamefont {{Viel}}},\ and\ \bibinfo {author} {\bibfnamefont {L.}~\bibnamefont {{Verde}}},\ }\bibfield  {title} {\bibinfo {title} {{Minimally parametric power spectrum reconstruction from the Lyman {\ensuremath{\alpha}} forest}},\ }\href {https://doi.org/10.1111/j.1365-2966.2011.18245.x} {\bibfield  {journal} {\bibinfo  {journal} {\mnras}\ }\textbf {\bibinfo {volume} {413}},\ \bibinfo {pages} {1717} (\bibinfo {year} {2011})},\ \Eprint {https://arxiv.org/abs/1010.1519} {arXiv:1010.1519 [astro-ph.CO]} \BibitemShut {NoStop}%
\bibitem [{\citenamefont {Inomata}\ \emph {et~al.}(2016)\citenamefont {Inomata}, \citenamefont {Kawasaki},\ and\ \citenamefont {Tada}}]{Inomata:2016uip}%
  \BibitemOpen
  \bibfield  {author} {\bibinfo {author} {\bibfnamefont {K.}~\bibnamefont {Inomata}}, \bibinfo {author} {\bibfnamefont {M.}~\bibnamefont {Kawasaki}},\ and\ \bibinfo {author} {\bibfnamefont {Y.}~\bibnamefont {Tada}},\ }\bibfield  {title} {\bibinfo {title} {{Revisiting constraints on small scale perturbations from big-bang nucleosynthesis}},\ }\href {https://doi.org/10.1103/PhysRevD.94.043527} {\bibfield  {journal} {\bibinfo  {journal} {Phys. Rev. D}\ }\textbf {\bibinfo {volume} {94}},\ \bibinfo {pages} {043527} (\bibinfo {year} {2016})},\ \Eprint {https://arxiv.org/abs/1605.04646} {arXiv:1605.04646 [astro-ph.CO]} \BibitemShut {NoStop}%
\bibitem [{\citenamefont {Graham}\ and\ \citenamefont {Ramani}(2024)}]{Graham:2024hah}%
  \BibitemOpen
  \bibfield  {author} {\bibinfo {author} {\bibfnamefont {P.~W.}\ \bibnamefont {Graham}}\ and\ \bibinfo {author} {\bibfnamefont {H.}~\bibnamefont {Ramani}},\ }\bibfield  {title} {\bibinfo {title} {{Constraints on dark matter from dynamical heating of stars in ultrafaint dwarfs. II. Substructure and the primordial power spectrum}},\ }\href {https://doi.org/10.1103/PhysRevD.110.075012} {\bibfield  {journal} {\bibinfo  {journal} {Phys. Rev. D}\ }\textbf {\bibinfo {volume} {110}},\ \bibinfo {pages} {075012} (\bibinfo {year} {2024})},\ \Eprint {https://arxiv.org/abs/2404.01378} {arXiv:2404.01378 [hep-ph]} \BibitemShut {NoStop}%
\bibitem [{\citenamefont {Font}\ \emph {et~al.}(2001)\citenamefont {Font}, \citenamefont {Navarro}, \citenamefont {Stadel},\ and\ \citenamefont {Quinn}}]{Font:2001py}%
  \BibitemOpen
  \bibfield  {author} {\bibinfo {author} {\bibfnamefont {A.~S.}\ \bibnamefont {Font}}, \bibinfo {author} {\bibfnamefont {J.~F.}\ \bibnamefont {Navarro}}, \bibinfo {author} {\bibfnamefont {J.}~\bibnamefont {Stadel}},\ and\ \bibinfo {author} {\bibfnamefont {T.~R.}\ \bibnamefont {Quinn}},\ }\bibfield  {title} {\bibinfo {title} {{Halo substructure and disk heating in a lambda cdm universe}},\ }\href {https://doi.org/10.1086/338479} {\bibfield  {journal} {\bibinfo  {journal} {Astrophys. J. Lett.}\ }\textbf {\bibinfo {volume} {563}},\ \bibinfo {pages} {L1} (\bibinfo {year} {2001})},\ \Eprint {https://arxiv.org/abs/astro-ph/0106268} {arXiv:astro-ph/0106268} \BibitemShut {NoStop}%
\end{thebibliography}%

\end{document}